\documentclass[preprint]{aastex701}

\usepackage{graphicx}

\shorttitle{Centaur Nuclei: Sizes, Shapes, Spins, and Structure}
\shortauthors{Fernandez, Buie, Lacerda, and Marschall}

\begin{document}


\title{Centaur Nuclei: Sizes, Shapes, Spins, and Structure\footnote{This
is an adapted, preprint version of the document that appears as
Chapter 4 in the book {\sl Centaurs} (K. Volk, M. Womack, J. Steckloff, Eds.), 
IOP Publishing, Bristol, UK, pp. 4-1 to 4-27, 
DOI 10.1088/2514-3433/ada267ch4 (2025).}}

\author[orcid=0000-0003-1156-9721]{Y. R. Fern\'andez}
\affiliation{Department of Physics, University of Central Florida, Orlando, USA}
\email[show]{yan@ucf.edu}  

\author[orcid=0000-0003-0854-745X]{M. W. Buie} 
\affiliation{Southwest Research Institute, Boulder, USA}
\email{buie@boulder.swri.edu}

\author[orcid=0000-0002-1708-4656]{P. Lacerda}
\affiliation{Instituto de Astrof\'\i sica e Ci\^ encias do Espa\c co, 
Universidade de Coimbra, Coimbra, Portugal}
\affiliation{Instituto Pedro Nunes, Coimbra, Portugal}
\email{lacerda.pedro@gmail.com}

\author[orcid=0000-0002-0362-0403]{R. Marschall}
\affiliation{Centre National de la Recherches Scientifique, 
Observatoire de la C\^ote d’Azur, Nice, France}
\email{raphael.marschall@oca.eu}


\begin{abstract}

We present a wide-ranging but in-depth analysis 
of Centaurs, focusing on their physical and structural aspects. 
Centaurs, originating from the Scattered Disk and Kuiper Belt, 
play a crucial role in our understanding of Solar System evolution. 
We first examine how biases in discovery and measurement affect 
our understanding of the Centaur size distribution. In particular 
we address the strong dependence of the census on perihelion 
distance and the broad distribution of Centaur geometric albedos. 
We explore the rotational characteristics derived from lightcurves, 
revealing a diverse range of spin rates and photometric 
variabilities, with most Centaurs showing low amplitude lightcurves,
suggesting near-spherical shapes. Additionally, we investigate the
relationships between Centaur orbital parameters, surface colors, 
and physical properties, noting a lack of correlation between 
rotational dynamics and orbital evolution. We also address the 
influence of sublimation-driven activity on Centaur spin states, 
and the rarity of contact binaries. We then discuss some observational 
and modeling limitations from using common observations (e.g. 
visible or infrared photometry) to determine diameters and shapes. 
Following that, we give some points on understanding how Centaur 
diameters and shapes can reveal the `primitive' nature of the 
bodies, emphasizing the important role occultation observations play. 
We also then assess how the Centaur size distribution we see 
today has been influenced by the collisions in both the primordial 
Kuiper Belt and in the subsequent Scattered Disk.  Finally, we end 
the chapter with a short narrative of future prospects for overcoming 
our current limitations in understanding Centaur origins and evolution.

\end{abstract}




\section{Introduction}

As described by other works in this
volume 
\citep[e.g., Chapters 1 and 7,][]{steckloffchap1,kokotanekovachap7},
the current sizes, shapes, spin states, and structures of Centaurs
are manifestations of the evolutionary processes that they
have suffered since the original icy planetesimals were first accreted
\citep[see, e.g., Chapter 2,][]{johansenchap2}.
In principle, if we were to know these physical properties
of the entire Centaur population, and of the small-body groups that
the Centaurs are related to, then we would have useful constraints
that would have to be matched by any model purporting to explain
the evolution.

Centaurs in particular play a unique role in our quest to
understand Solar System evolution, since they come from
the Scattered Disk (i.e., as Scattered Disk Objects, SDOs) 
and the Kuiper Belt,
and since many of them end up as 
Jupiter-family comets (JFCs);
see, e.g., \cite{Duncan1997Sci,2004come.book..193D,2008ApJ...687..714V}
and Chapter 3 \citep{disistochap3}. 
A broad objective with studying
Centaur physical properties is thus to contextualize what
we see among the SDOs and JFCs. 
Part of this objective 
is an understanding of just if and how the
Centaurs change before and during the dynamical cascade from the
Scattered Disk, and
what happens to the Centaurs in the (typically) few $10^6$ yr 
of their residence in the giant-planet region. 

While there is much ongoing work to understand the
physical properties of Centaurs, a murky fog 
hinders us in several ways. 
First, as with many groups of small bodies in the
Solar System, the characterization of Centaurs lags somewhat
behind their discovery. There are many Centaurs with only 
minimal characterization. Second, 
the discovery of Centaurs itself has biases, so even if we were
characterizing all known Centaurs well, there would still be
gaps in our knowledge because we are simply ignorant of many
of the faintest and smallest objects in the population. 
Third, the techniques that we use to measure Centaur
brightnesses and thereby extract physical properties
have their own set of limitations. All of these
problems conspire to bias our picture of what we know
about Centaur ensemble properties, and eventually
they must be all well understood if we are to properly
interpret our data. 

In this chapter we focus on two themes that touch
on these issues. 
(1) What are the current constraints on Centaur 
physical properties, and how might limitations in
our measurement techniques be biasing those results?
(2) What is the context of Centaurs in terms of
primitive-body evolution in the Solar System?


\section{Size Distribution}

\begin{figure}
\begin{center}
\includegraphics[width=0.9\textwidth]{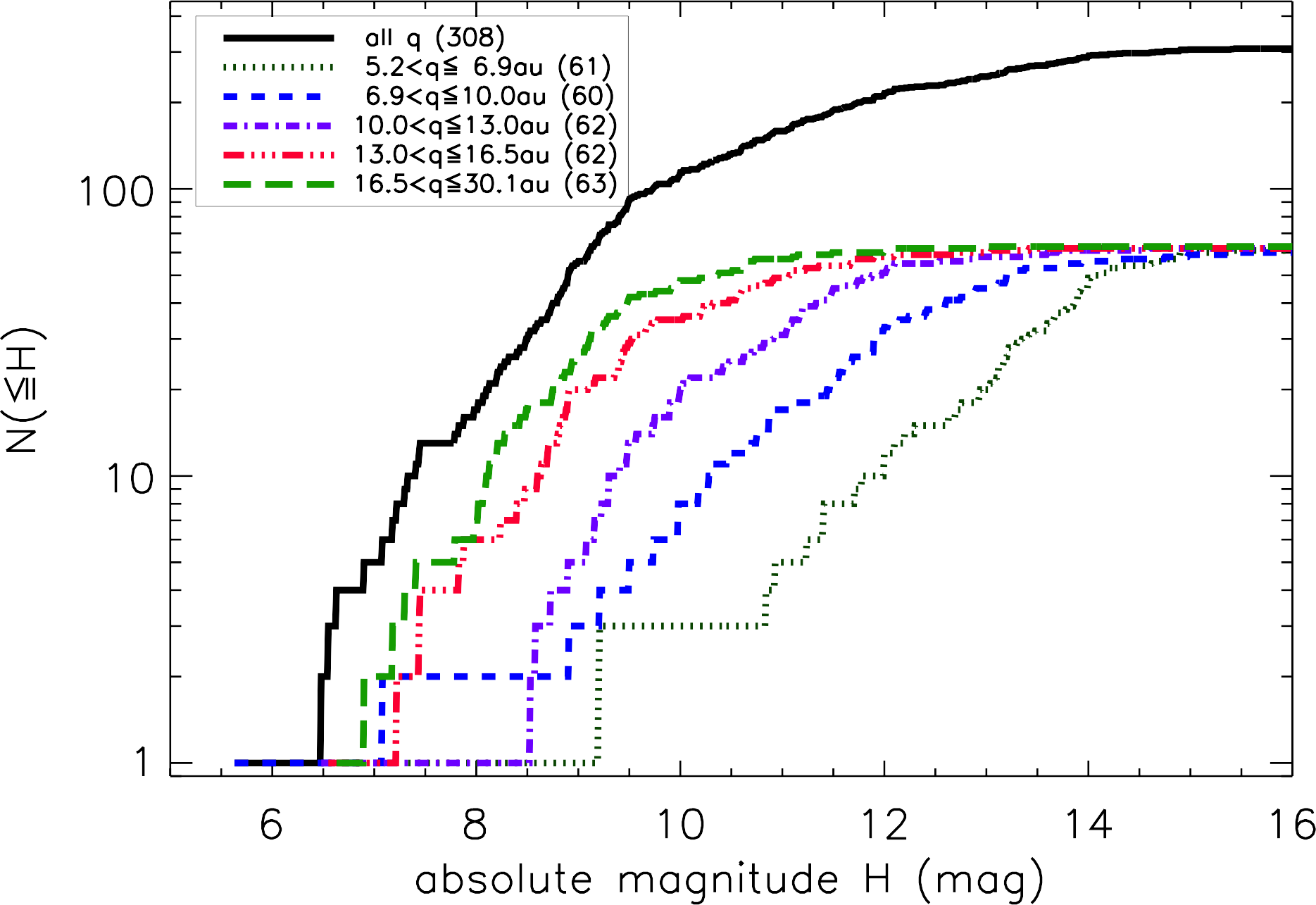}
\end{center}
\caption{Cumulative magnitude
distribution (CMD) of known asteroidal 
Centaurs. All objects (as described in the text) are in
the thick black line. The other five thick linestyles with color show
CMDs for various perihelion ($q$) bins, tuned so that each bin
has approximately the same number of Centaurs. There are very strong
differences in the five curves indicating that the completion of the
Centaur census does not extend to very faint magnitudes yet.\label{fig:magdist}}
\end{figure}

Much of the physics of this topic is discussed in
Chapter 7
\citep{kokotanekovachap7}, so we describe a few
statistical points here. 
A very basic look at the current absolute magnitude ($H$)
distribution of the known Centaurs is shown in
Figure~\ref{fig:magdist}. We extracted all
asteroid entries in the JPL Horizons database
(as of 2023 November 21) that had
$q>5.203$ au (i.e., Jupiter's orbital semimajor axis $a_J$) 
and $a<30.1$ au (i.e., Neptune's orbital semimajor axis $a_N$), 
but excluding objects that are (or could be) from
different source regions: 
Jovian and Neptunian Trojans,
parabolic and hyperbolic objects, and
nine additional objects with inclinations above 
$80^\circ$. 
This left 308 Centaurs. We note that 
this $a_N$ cutoff in the Centaur definition
does introduce some bias in the distant Centaurs since there 
is no real physical/dynamical break here between
Scattered Disk and Centaur region; e.g.
(523746) 2014 UT$_{114}$ with $a=30.09$ au makes our cut,
but (309239) 2007 RW$_{10}$ with $a=30.16$ au
does not. We also note that even if the
high-inclination/retrograde Centaurs are indeed
from the Scattered Disk anyway \citep[see, e.g.,][]{2020MNRAS.494.2191N}
this is a small
number and does not affect the point we are making here.

The magnitude distribution of those 308 
Centaurs are represented by the thick solid line in the
Figure~\ref{fig:magdist}.
The curves with other linestyles/colors show
the distribution broken up into perihelion
bins, as listed in the figure's legend. 
We decided to create bins
where approximately equal number of Centaurs 
would be in each one. 
The idea is to show that the discovery
rates very strongly depend on how close the
Centaurs come to our telescopes here in the inner Solar System. For
the smallest perihelion bin (green dotted line),
the magnitude distribution seems to be a single power law
that turns over due to lack of detected objects
around $H=14$. In contrast, the other curves
are less straight and of course also turn over at brighter 
values of $H$. The most distant bin, the light green
dashed line, has a turn over about 5 full magnitudes
brighter than the closest bin. One could argue
from this that we are only really confident in our
census of the Centaurs down to about $H=9$, and we 
certainly have
not yet discovered enough Centaurs in
the $H=9$ to $14$ range to permit characterizing the population
in this size range without large corrections due to discovery biases.

\begin{figure}
\begin{center}
\includegraphics[width=0.9\textwidth]{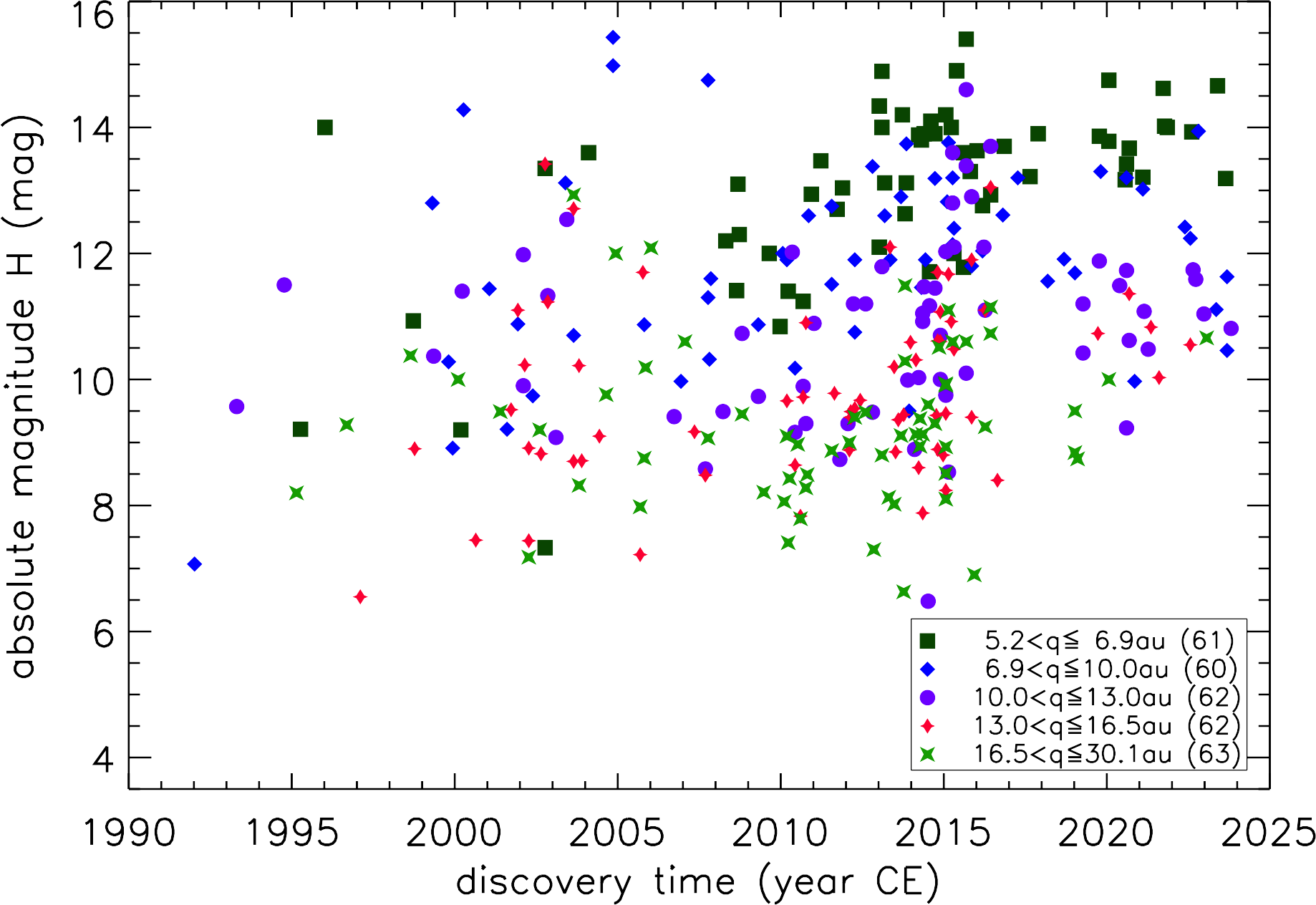}
\end{center}
\caption{Scatter plot of when asteroidal Centaurs
of various $q$ and $H$ were discovered. Color coding
and perihelion binning
match that in Figure~\ref{fig:magdist}.
\label{fig:discyear}}
\end{figure}

Another demonstration of the aforementioned perihelion 
effect is
shown in Figure~\ref{fig:discyear}, which
plots the discovery time versus a Centaur's
$H$. The color coding here is the same
as in the previous figure. 
For the Centaurs that are presumably
easier to discover, i.e. those in the closest perihelion
bin (green squares), there is some indication
that we have discovered all or nearly all of
the relatively large Centaurs ($H\approx13$ and brighter). 
In the last
few years, the only such Centaurs discovered
in that bin are around $H=13$ to 15. This is
true even though clearly the sky surveys are
still capable of finding brighter Centaurs,
since those are in fact being found in the
more distant perihelion bins. Interestingly,
no Centaur brighter than about $H=8.7$ has been 
discovered since late 2016 (7 years ago at time of writing). 

\begin{figure}
\begin{center}
\includegraphics[width=0.9\textwidth]{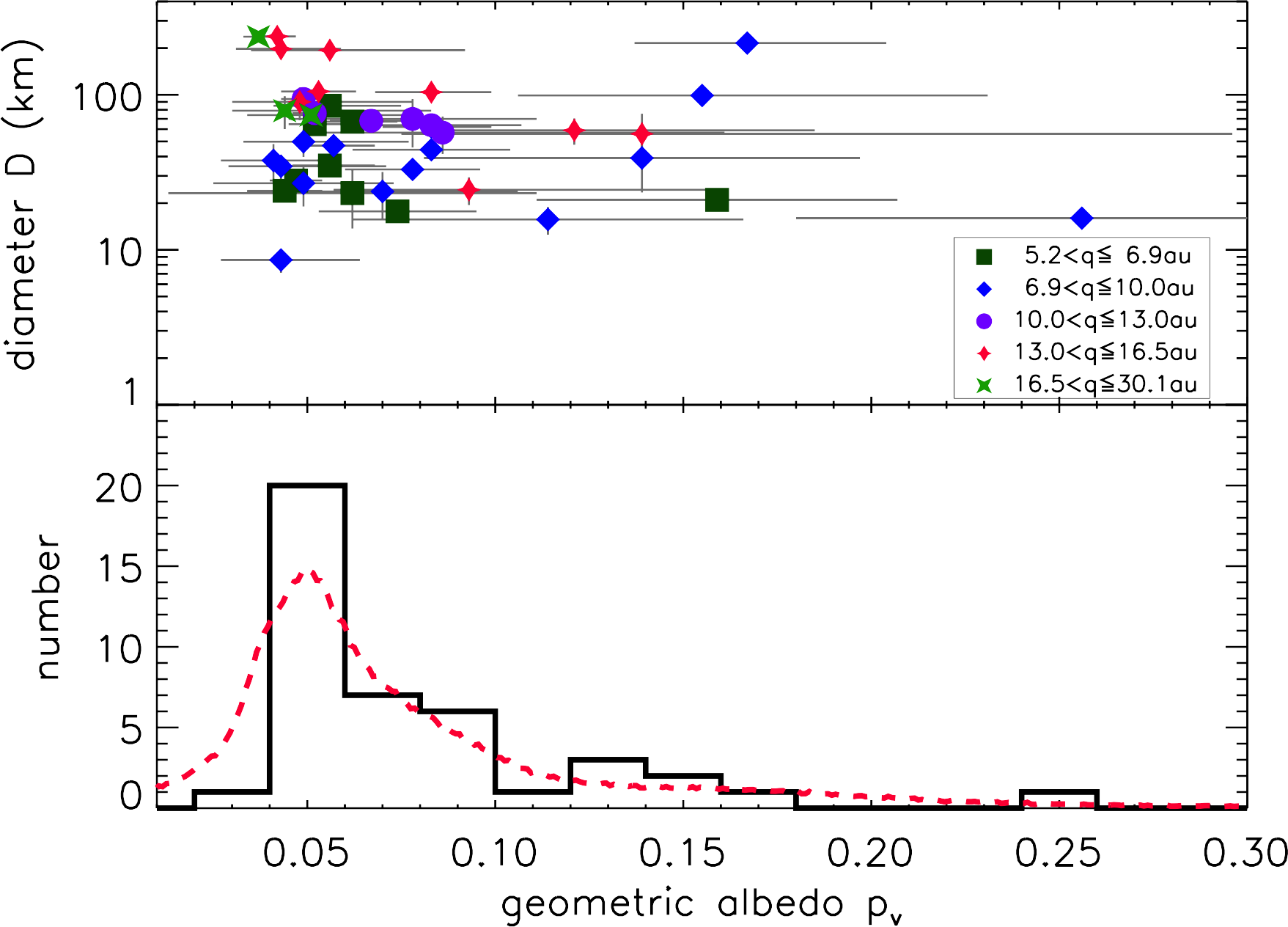}
\end{center}
\caption{Top: Scatter plot of 42 Centaur diameters $D$
and geometric albedos $p_v$ as summarized by \cite{2020tnss.book..153M}. Color coding matches
that in Figures~\ref{fig:magdist} and \ref{fig:discyear}. Bottom: 
Histogram of those 42 albedos. The red dashed curve
is our estimate of the albedo probability density
function (PDF) based on the given albedos. 
We estimated the PDF by generating simulated albedos 
based on the measured values and their uncertainties.
\label{fig:albs}}
\end{figure}

The fundamental property that we wish to 
understand is the size distribution, not the
magnitude distribution, and so the geometric
albedo is required to make the conversion.
(See, e.g., \cite{2007Icar..190..250P} for a definition
of this albedo.)
With JFCs, the spread of albedos seems to be 
fairly limited \cite[see, e.g.,][]{2004come.book..223L}, but
this is not the case for Centaurs. Using a
recent compilation of outer Solar System albedos by
\cite{2020tnss.book..153M}, we make use
of 42 Centaur albedos in their Table 7.1. 
(The table itself lists 55 Centaurs, 
but here we exclude 3 active comets for
which the nucleus albedo may be difficult to get
(29P, 167P, and C/2011 KP36) as well as 10 other
objects that \cite{2020tnss.book..153M} included
as Centaurs even though their perihelia are 
smaller than $a_J$.) 
A scatter plot of those 42 albedos is shown in 
the top of Figure~\ref{fig:albs}.
The color coding for perihelion bins 
is again the same as the previous figures. There is
no trend of albedo with diameter, and no significant
difference between the albedos in any perihelion
bin (via a check of the Kolmogorov-Smirnov (KS) test). 
At the moment, if one were to assume that albedo is 
controlled by the surface evolution of the Centaur
\citep[see Chapter 5, Chapter 6, Chapter 7, Chapter 8;][]{peixinhochap5,mandtchap6,kokotanekovachap7,bauerchap8},
there is apparently no indication that surfaces of 
Centaurs with smaller $q$ 
have evolved differently than those of the Centaurs
farther out. This might be simply a manifestation
of the fact that Centaurs do  
jostle toward and away from the Sun in the 
Centaur region, even if there is an overall
trend of the Centaurs (at least the ones
that survive to become JFCs) slowly moving 
inward \citep{2004come.book..193D,2024come.book..121F}.

Given the absence of a strong
trend of albedo with perihelion or semimajor axis,
we will take the current albedo distribution
for those 42 Centaurs as being representative of
the population as a whole, and use it as a
probability density function (PDF) in order to 
explore what the size distribution might be
as derived from the $H$ distribution. 
In other words, since we have albedos for less than 
$1/7$ of the
Centaurs (42 out of 308), the conversion requires
assuming an albedo distribution.
The histogram of those 42 albedos is shown in 
the bottom panel of Figure~\ref{fig:albs}. 
We estimated a PDF from those 42 measurements
and their error-bars by simulating 10,000 albedo
distributions Monte Carlo-style. The result is
the red dashed
curve in the bottom plot of Figure~\ref{fig:albs}. 
Note that this is not a fit to the albedo
distribution, simply an empirical assessment
of what the true underlying distribution might be.

\begin{figure}
\begin{center}
\includegraphics[width=0.9\textwidth]{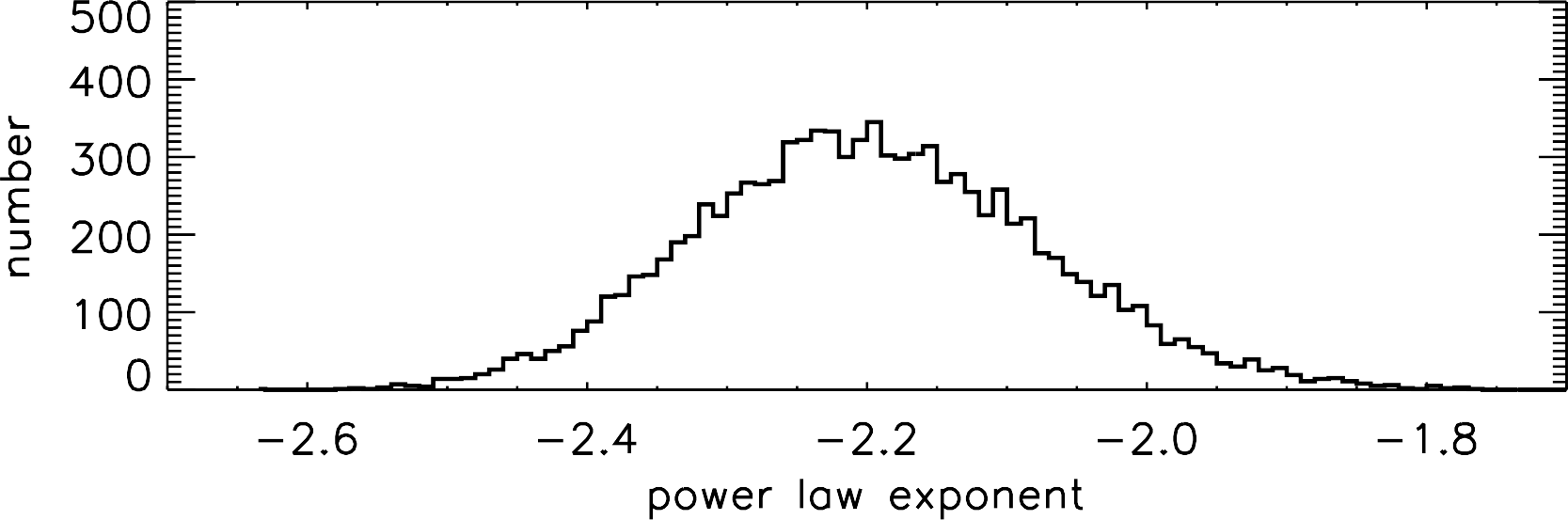}
\end{center}
\caption{Distribution of power-law slopes 
from our simulation of 10000 cumulative size distributions
created as described in the text. While the 
most likely result of our particular
fit scheme is around $-2.2$, 
there is a wide range of other possible exponents, and
this 
must be taken into account when interpreting
the Centaur size distribution. Note that we are
not claiming a value for the actual Centaur size
distribution, we are merely demonstrating that 
whatever power-law is produced has appreciable 
uncertainty.
\label{fig:powlaw}}
\end{figure}

Using this PDF, we then simply ran 10,000 simulations that
each created a size distribution from the
magnitude distribution by randomly extracting
an albedo for each object according to 
the probability density. We then fit a power-law
to each cumulative size distribution with
diameters between 100 and 200 km. This range was chosen
as one for which the
distribution seems to follow a single power-law; i.e.
it made the fitting seem sensible. 
The distribution
of those 10000 power-law exponents is shown
in Figure~\ref{fig:powlaw}. 
While there is a most-likely result near a power-law
exponent 
of $-2.2$, there
is a reasonable probability of getting an
exponent that is well off from
that answer. We emphasize that
the main point here is not the
actual value of the exponent, but rather that
until a higher fraction of the Centaur
diameters and albedos (not just their $H$ magnitudes)
have actually been measured, 
there is intrinsic uncertainty
in any characterization of the
Centaur size distribution. This has
ramifications for any attempt to 
use a physical model to explain such a size distribution or to
explain how Centaurs fit into the SDO and JFC size distributions.
This is above and beyond other problems that could
make such hypothesizing difficult -- e.g. the fact that
there is almost no overlap between the known JFC diameters
and the known Centaur diameters.

It can be argued that the size distribution is not actually
the most fundamental property we might use to gauge formation
and evolutionary processes. Rather in some contexts
the mass distribution might be more important. This is 
currently even more fraught than the size assessment is
since to go from size to mass requires a density, and
we have precious little info on the densities, porosities, and overall
structure of the Centaurs
\citep[see \ref{sec:densityconstraints} below, and Chapters 9 and 12;][]{sickafoosechap9,hirabayashichap12}.


\section{Shape and Spin Distributions}

\label{sec:shpspin}

Determining the shapes and spins of Centaurs involves studying their lightcurves, which record periodic oscillations in brightness over time. Well sampled lightcurves are processed to extract rotational properties: the period, $P$, and the full range of photometric variation, $\Delta m$. The former corresponds to the spin period, while the latter constrains the shape of individual objects. If measured at different observing geometries, time series data can be inverted to extract more precise shapes and spin pole orientations. In some cases, under favorable observing geometries, lightcurves can also reveal whether the object has multiple components (bilobed shape as the nucleus 67P, or two separate components as KBO Arrokoth).  The distributions of rotational properties are useful for comparing different populations when trying to decode evolutionary links between them.

Other techniques exist to decipher the shapes of Centaurs. For bright objects, adaptive optics systems on ground-based telescopes can also be used to estimate their shapes and dimensions. Stellar occultations are a powerful technique to constrain shapes, capable to achieving extraordinary resolution on chords across the sky-projected cross-section of individual objects. These techniques are often complementary.

The Asteroid Lightcurve Database \citep{2009Icar..202..134Warner} currently includes lightcurves properties, spin period and photometric range, for 16 Centaurs. Of those, 7 have been observed at more than one observing geometry, revealing changing variability. Table \ref{tab:rotational_properties} summarises the known Centaur rotational properties, listed together with other orbital, physical and surface parameters. Figures~\ref{fig:spinfreq_distr}-\ref{fig:obliquity_jacobi_model_no_coma} highlight a few aspects of data discussed in this chapter. 

\begin{deluxetable}{lrrrrrrrrrrl}
\tabletypesize{\footnotesize}
\tablewidth{0pt} 
\tablecaption{Known Centaur Rotational Properties\label{tab:rotational_properties}}
\tablehead{
\colhead{} & \colhead{} & \colhead{} & \colhead{} & \colhead{} & \colhead{} & 
    \colhead{} & \colhead{Abs.} & \colhead{} & \colhead{} & \colhead{} & \colhead{} \\
\colhead{} & \colhead{Lightcurve} & \colhead{} & \colhead{} & \colhead{} & \colhead{} & 
    \colhead{} & \colhead{Mag.} & \colhead{$B-R$} & \colhead{Surface} & \colhead{} & \colhead{} \\
\colhead{} & \colhead{Period\tablenotemark{a}} & \colhead{Min. $\Delta m$} & \colhead{Max. $\Delta m$} & \colhead{$q$} & \colhead{} & 
    \colhead{$i$} & \colhead{$H$} & \colhead{Color} & \colhead{Geometric} & \colhead{Diameter} & \colhead{} \\
\colhead{Centaur} & \colhead{[hr]} & \colhead{[mag]} & \colhead{[mag]} & \colhead{[au]} & \colhead{$e$} & 
    \colhead{[deg]} & \colhead{[hr]} & \colhead{[mag]} & \colhead{Albedo} & \colhead{$D$ [km]} & \colhead{Notes\tablenotemark{b}} 
}
\startdata
2060 Chiron (1977 UB)       & 5.92   & 0.04                    & 0.09                    & 8.55  & 0.376 & 6.9  & 6.56  & 1.01  & $0.16\pm 0.03$ & $218	\pm	20$ & U3, C, R?\\
5145 Pholus (1992 AD)       & 9.98   & 0.15                    & 0.6                     & 8.67  & 0.573 & 24.7 & 7.64  & 1.97  & $0.16\pm 0.08$ & $99	\pm	15$ & U3       \\
8405 Asbolus (1995 GO)      & 8.94   & 0.32                    & 0.55                    & 6.88  & 0.619 & 17.6 & 9.19  & 1.228 & $0.06\pm 0.02$ & $85	\pm	9$  & U3       \\
10199 Chariklo (1997 CU26)  & 7.00   &                         & 0.11                    & 13.14 & 0.168 & 23.4 & 6.65  & 1.299 & $0.04\pm 0.01$ & $248	\pm	18$ & U2, R\\
31824 Elatus (1999 UG5)     & 26.82  &                         & 0.1                     & 7.22  & 0.386 & 5.3  & 10.42 & 1.672 & $0.05\pm 0.03$ & $50	\pm	10$ & U2       \\
32532 Thereus (2001 PT13)   & 8.34   & 0.16                    & 0.38                    & 8.50  & 0.199 & 20.4 & 9.29  & 1.19  & $0.07\pm 0.02$ & $74	\pm	17$ & U3       \\
52872 Okyrhoe (1998 SG35)   & 9.72   & 0.07                    & 0.4                     & 5.82  & 0.305 & 15.7 & 11.23 & 1.237 & $0.06\pm 0.02$ & $36\pm	1$  & 12.17,U2 \\
54598 Bienor (2000 QC243)   & 9.14   & 0.08                    & 0.75                    & 13.20 & 0.203 & 20.7 & 7.69  & 1.158 & $0.04\pm 0.02$ & $198	\pm	7$  & U3       \\
60558 Echeclus (2000 EC98)  & 26.80  &                         & 0.24                    & 5.84  & 0.457 & 4.3  & 9.55  & 1.376 & $0.05\pm 0.02$ & $65	\pm	2$  & U2, C\\
83982 Crantor (2002 GO9)    & 13.94  &                         & 0.14                    & 14.1  & 0.274 & 12.8 & 9.17  & 1.864 & $0.12\pm 0.06$ & $59	\pm	12$ & 19.34,U1 \\
95626 (2002 GZ32)           & 5.8    &                         & 0.08                    & 18.01 & 0.219 & 15.0 & 7.39  & 1.199 & $0.04\pm 0.01$ & $237	\pm	8$  & U3       \\
136204 (2003 WL7)           & 8.24   &                         & 0.05                    & 14.92 & 0.258 & 11.2 & 8.73  & 1.23  & $0.05\pm 0.01$ & $105	\pm	7$  & S,U2-    \\
145486 (2005 UJ438)         & 8.32   &                         & 0.13                    & 8.22  & 0.532 & 3.8  & 10.89 & 1.64  & $0.24\pm 0.12$ & $16	\pm	2$  & A,U2-    \\
250112 (2002 KY14)          & 8.50   & 0.09                    & 0.13                    & 8.62  & 0.314 & 19.5 & 9.74  & 1.75  & $0.12\pm 0.09$ & $43	\pm	6$  & U2       \\
459865 (2013 XZ8)           & 87.74  &                         & 0.26                    & 8.42  & 0.369 & 22.6 & 9.53  & 1.17  &                & $69	\pm	35$\tablenotemark{c} & U2       \\
471931 (2013 PH44)          & 22.16  &                         & 0.15                    & 15.57 & 0.213 & 33.5 & 9.38  &       &                & $74	\pm	37$\tablenotemark{c} & A,U1  \\  
\enddata
\tablecomments{Lightcurve properties taken from Lightcurve Database \citep{2009Icar..202..134Warner}, updated October 2023. Diameters and albedos come from Herschel \citep{2013AandA...555A..15Fornasier, 2014AandA...564A..92Duffard}, or NEOWISE \citep{2019PDSS..251.....Mainzer} surveys; the least uncertain of the two measurements is taken when they agree within 1$\sigma$, otherwise the mean is shown. Colours taken from \cite{2012AandA...546A..86Peixinho} and \cite{2016AJ....152..210Tegler}.}
\tablenotetext{a}{Lightcurve period, where all are double peaked except where indicated in notes.}
\tablenotetext{b}{``U$x$(-)'' is the quality of the lightcurve from $x=1$ (worse) to $x=3$ (best) and ``-'' is a half-step, ``S'' indicates a single-peaked period is given, ``A'' indicates ambiguity between double-peaked and single-peaked period, and a number indicates another possible period. ``C'' indicates cometary activity, ``R'' means rings detected (followed by ``?'' if unconfirmed).}
\tablenotetext{c}{An albedo of 0.057 is assumed to calculate diameter from $H$ for last two rows.}
\end{deluxetable}

\subsection{Period and $\Delta m$ distributions}

\begin{figure}
\begin{center}
\includegraphics[width=0.9\textwidth]{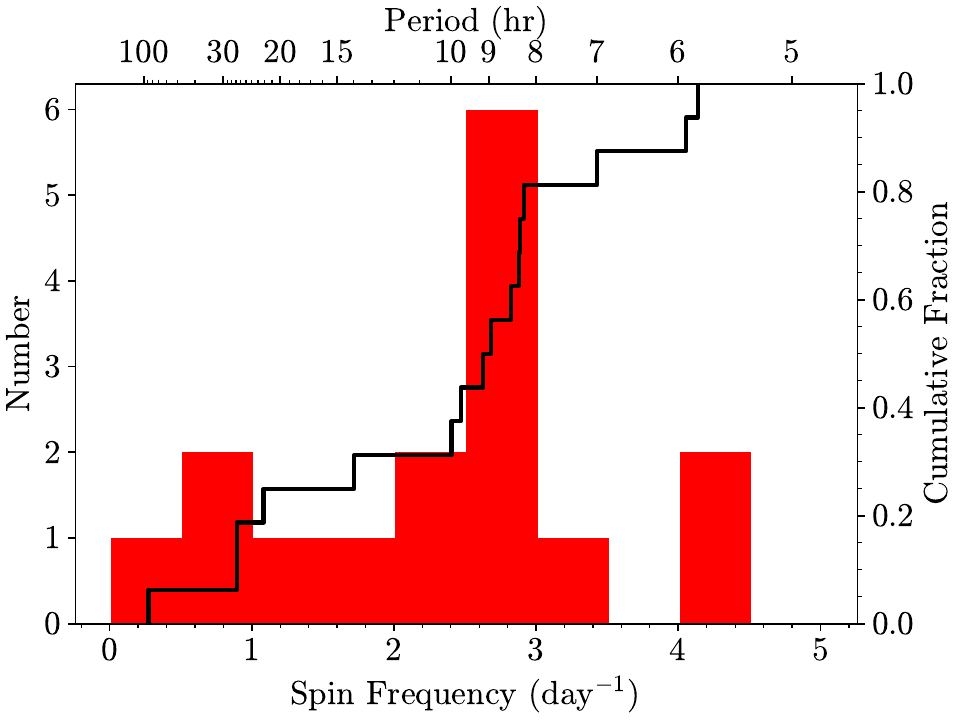}
\end{center}
\caption{Centaur spin rate distribution. Cumulative distribution (solid black line) is plotted on the right vertical axis.\label{fig:spinfreq_distr}}
\end{figure}

\begin{figure}
\begin{center}
\includegraphics[width=0.9\textwidth]{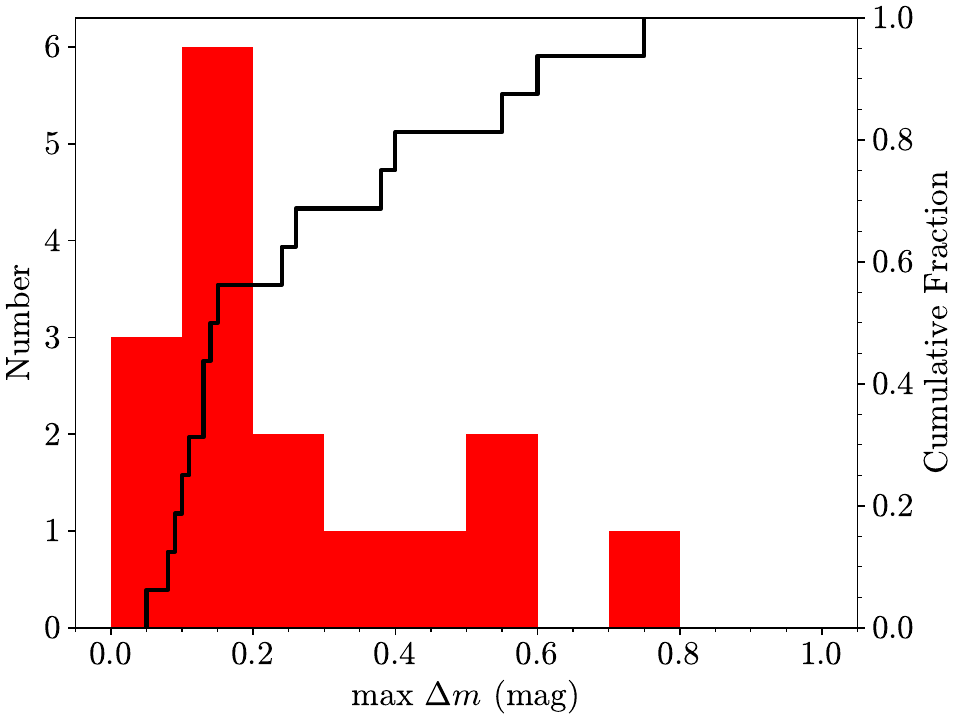}
\end{center}
\caption{Distribution of maximum lightcurve variability for Centaurs. Cumulative distribution (solid black line) is plotted on the right vertical axis. \label{fig:dm_distr}}
\end{figure}

Figures~\ref{fig:spinfreq_distr} and \ref{fig:dm_distr} show histograms and cumulative distributions of spin frequency and lightcurve variability. Even though the period is often easier to grasp, frequency is the more relevant physical quantity; we will refer to both interchangeably. Six of the 16 Centaurs cluster between 2.5 and 3 rotations per day ($P\sim 9$ hr) near the center of the distribution. A long tail extends to much slower rotations, with 3 objects spinning slower than $P=24$ hr and Centaur 2013 XZ8 spinning with $P\sim 88$ hr. The two fastest rotators are Chiron and 2002 GZ32 with periods near 6 hr.

The $\Delta m$ distribution is skewed towards low values, with roughly half the Centaurs within $\Delta m<0.2$ mag. At the opposite end of the distribution, Bienor is the most variable object at $\Delta m=0.75$ mag. As of yet no Centaurs display the extreme variability ($\Delta m>0.9$ mag) characteristic of extreme bilobed or compact binary shapes (although it is possible that there is a bias against discovering such objects, since perhaps the large lightcurve amplitude means it is less likely for such objects to be seen at follow-up  observations right after discovery). We discuss binarity in more detail below. Another bias against discovery highly variable objects is that larger objects, which are easier to detect, will tend to be more spherical (although high-$\Delta m$ Bienor is the 4th largest object in Table \ref{tab:rotational_properties}.)

\begin{figure}
\begin{center}
\includegraphics[width=0.9\textwidth]{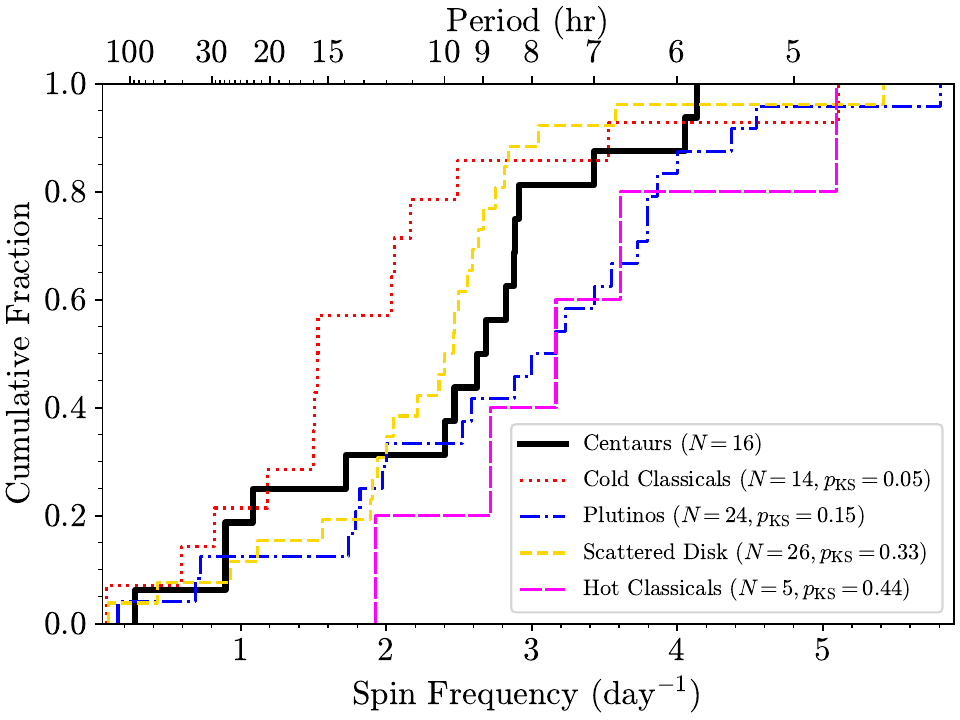}
\end{center}
\caption{Cumulative spin rate distributions of Centaurs and other transneptunian objects. Legend indicates, for each dynamical group, sample size and the Kolmogorov-Smirnov test $p$-value that the sample and the Centaurs have the same spin rate distribution.\label{fig:spinfreq_distr_comparison}}
\end{figure}

Figure~\ref{fig:spinfreq_distr_comparison} compares the spin rate distribution of Centaurs with different dynamical families of transneptunian objects. We use the two-sample KS test\footnote{We used the implementation provided in Python module {\tt scipy.stats.ks\_2samp}.} to compare the Centaurs with the other classes of objects. Cold Classical have the least compatible spin distribution with the Centaurs ($p_\mathrm{KS}$-value of 0.05). The KS-test does not rule that Centaur spins come from the same distribution as Plutinos ($p_\mathrm{KS}=0.15$), Scattered Disk objects ($p_\mathrm{KS}=0.33$) or Hot Classicals ($p_\mathrm{KS}=0.44$). However, the sample sizes are small so more data is needed before strong conclusions can be made. A similar exercise comparing the max $\Delta m$ distributions of Centaurs and the same transneptunian dynamical families shows no discernible differences, with $p_\mathrm{KS}>0.66$ in all cases.

\subsection{Relation to orbits}

Centaurs display no obvious relation between orbital parameters (perihelion distance, orbital eccentricity and inclination) and lightcurve properties. To zeroth order, this suggests that the processes responsible for the orbital evolution and the spin evolution are unrelated. One possible exception is that higher perihelion objects have lower $\Delta m$ lightcurves (although again there may be some bias against recovering such objects in the first place). Since Centaur activity correlates inversely with perihelion distance, this suggests a possible link between shape and activity. We discuss this possibility below. However, it would be interesting to study whether close approaches to giant planets can significantly modulate Centaur spins and shapes.

\subsection{Relation to surface properties}

The surface colors of Centaurs and small Kuiper Belt objects span a broad range from a solar $B-R\approx 1$~mag to a very red $B-R\approx 2$~mag. Furthermore, the color distribution appears bimodal, with a gap at $B-R\sim1.5$~mag \citep{2012AandA...546A..86Peixinho, 2020tnss.book..307Peixinho}. The reason for this surface color bimodality is unclear \citep{2020tnss.book..307Peixinho}. An initial explanation based on collisional resurfacing of radiation reddened surface with freshly excavated, neutrally colored material \citep{1996AJ....112.2310Luu} was ruled out \citep{1998ApJ...494L.117Luu,2003EM&P...92..233Thebault}. Cometary activity may also cause the resurfacing, affecting the colors of Centaurs and causing the bimodality \citep{2004A&A...417.1145Delsanti}. Indeed, all active Centaurs but one\footnote{The exception is 166P ($q=8.6$ au, $Q=19.2$ au, unknown rotational properties) which falls in neither the blue nor the red clump} occupy the bluer of the two color clumps. If low level, undetectable activity is present, this can also be due to intrinsically bluer coma dust diluting the redder color of the nucleus, or simply to nongeometric scattering in optically small particles dominating the coma and making it appear bluer \citep{2009AJ....137.4296Jewitt}. Another possibility is that the bimodality stems from a composition gradient with heliocentric distance in the planetesimal disk from which the Centaurs originate \citep[e.g.][]{2017AJ....153..145Wong,2019ApJ...880...71L}. 
Basically, depending on where a Centaur progenitor actually formed with respect to the condensation lines of various red species, a surface could evolve differently
due to the subsequent radiation and thermal
processing, perhaps leading to the discontinuity of colors that we see today. 

\begin{figure}
\begin{center}
\includegraphics[width=0.9\textwidth]{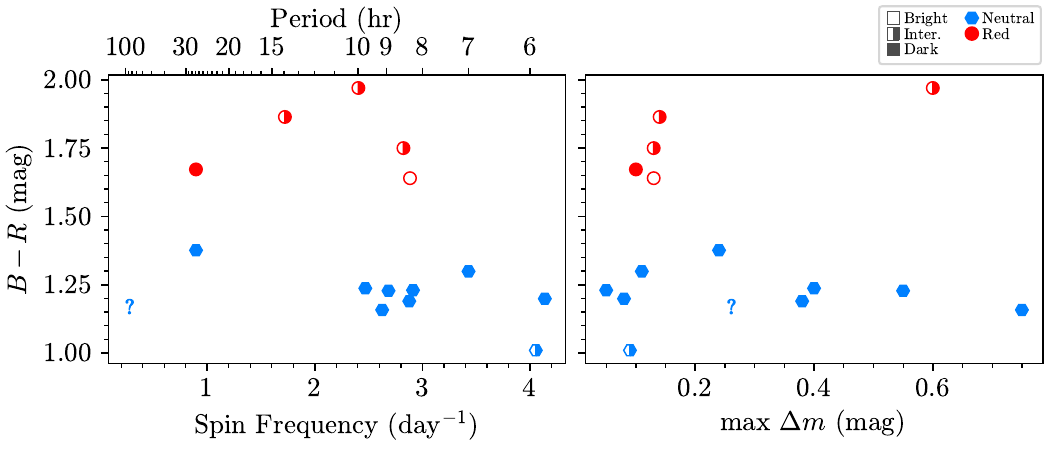}
\end{center}
\caption{Spin frequency (left panel) and maximum $\Delta m$ (right panel) versus $B-R$ color. Point color and shape highlight the bimodality and the fill pattern indicates surface albedo (filled symbols for ``Dark" albedos from 0.04 to 0.07; half-filled symbols for ``Intermediate" albedos from 0.12 to 0.16; open symbols for ``Bright" albedo 0.24; question mark indicates unknown albedo). \label{fig:rotprop_vs_color}}
\end{figure}

In any case these processes do not appear to correlate with the rotational properties of Centaurs, as shown in Figure~\ref{fig:rotprop_vs_color}. The different types of surfaces scatter across the entire range of observed spin frequencies. The red clump of Centaurs concentrates at low $\Delta m$, with the exception of Pholus, but this may be a small number fluctuation. An interesting observation not related to rotational properties is that blue Centaurs are darker. This is also seen in other outer Solar System populations \citep{2014ApJ...793L...2Lacerda}.

\subsection{Relation to Centaur size}

\begin{figure}
\begin{center}
\includegraphics[width=0.9\textwidth]{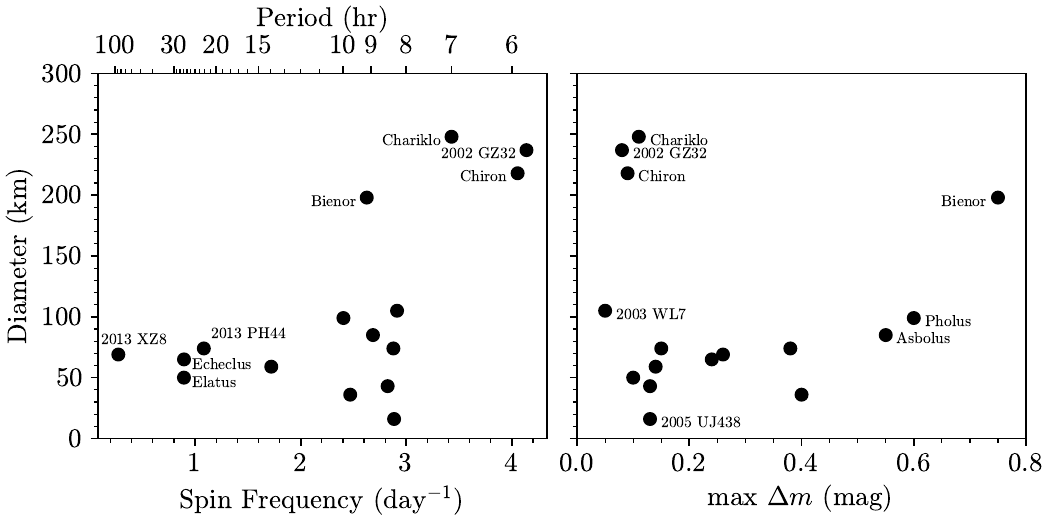}
\end{center}
\caption{Centaur diameter plotted against spin rate (left panel) and maximum lightcurve range (right panel). \label{fig:rotprop_vs_size}}
\end{figure}

Figure \ref{fig:rotprop_vs_size} (left panel) plots Centaur size against spin frequency and period. Smaller Centaurs ($D<100$ km) span a wide range of observed spin frequencies, from 0.5 to 3 rotations per day, with a median spin period of $\sim14$ hr. Larger Centaurs spin faster than 2.5 rotations per day, with a median period of $\sim7$ hr. The three largest Centaurs are also the fastest rotators, with $P<7$ hr. 

If a large proportion of the Centaurs are collisional fragments (see section \ref{sec:collisions}), then collisional evolution prior to leaving the source region must have played a role in setting rotations (see section~\ref{sec:processes_affecting_rotation}). However, energy equipartition would predict smaller Centaurs to spin faster, which argues against the observed distribution of spin frequencies being driven by collisional evolution. It may be that the smaller Centaurs hit a spin barrier at three rotations per day, which would suggest that they are gravitationally reaccumulated outcomes of collisions. If so, then the larger Centaurs were never disrupted by collisions, and retain their original spin rates. 

Another possible explanation for smaller Centaurs display slower spins has to do with torques due to activity. If spin-up and spin-down are equally likely, this tends to increase the median spin period as spun-up Centaurs may be rotationally disrupted leaving a population of survivors biased towards slower rotators. This effect, investigated in \cite{2021AJ....161..261Jewitt} for Jupiter-family comets (JFCs), is a function of size, with small objects more affected than larger ones: a factor of 10 increase in size corresponds to a factor 100 increase in spin-up/down timescale \citep[see also][]{2013ApJ...775L..10Samarasinha, 2018Icar..312..172Steckloff,2021PSJ.....2...14Safrit}. 

It is unlikely that activity can explain the presence of slower rotators and absence of faster rotators among smaller Centaurs relative to the spin rates of larger Centaurs. According to Fig.~1 in \cite{2021AJ....161..261Jewitt}, notional Centaur-sized nuclei ($D\sim100$ km) should have spin-up timescales $\sim 10^5$ yrs, comparable to the dynamical lifetimes of Centaurs. However, it is important to note that the Figure applies to JFCs with $q\sim 1$ to 2 au and hence larger mass-loss rates. An interesting complication is that some of the Centaurs may have dipped into temporary low-$q$ orbits () where activity could have played a more important role
\citep[Chapters 7 and 3;][]{kokotanekovachap7,disistochap3}. More work is needed to understand the effect of orbital evolution and activity on Centaurs spins.

\subsection{Shapes}

A simple way to translate lightcurve variability into a shape model is to assume that it is caused by a triaxial ellipsoid object rotating around its shortest principal axis. 
The caveat here is that in reality many objects are not ellipsoidal but have more complex shapes \citep[e.g.,][]{2019Sci...364.9771S,2018AJ....155..245B} making lightcurve interpretation more difficult.  Multiple lightcurves obtained at several orbital longitudes can help with this \citep{2001Icar..153...24K}, but that can be a lengthy proposition for a study of Centaurs. Nonetheless, we can continue with the simplifying assumption, and  the ellipsoid minimum and maximum sky-projected cross section areas are given by
\begin{equation}
    C_\mathrm{min}=\pi b \left(a^2 \cos^2\theta + c^2 \sin^2\theta\right)^{1/2}
\end{equation}
and
\begin{equation}
    C_\mathrm{max}=\pi a \left(b^2 \cos^2\theta + c^2 \sin^2\theta\right)^{1/2},
\end{equation}
\noindent where $a \geq b \geq c$ are the semi-axes of the ellipsoid, and $0\leq \theta\leq \pi/2$ is the aspect angle, measured between the line of sight and the spin pole direction.

The photometric range, $\Delta m$, is related to ratio of those areas and given by
\begin{equation}
  \Delta m = -2.5\log\frac{C_\mathrm{min}}{C_\mathrm{max}}=-1.25\log\left(\frac{\cos^2\theta + (c/a)^2\sin^2\theta}{\cos^2\theta + (c/b)^2\sin^2\theta}\right),
  \label{eq:photometric_range}
\end{equation}
\noindent which depends only on the ellipsoid axis ratios and the aspect angle. If the ellipsoid is observed equator-on ($\theta=\pi/2$), the axis ratio can be obtained directly from 
\begin{equation}
    b/a=10^{-0.4\Delta m}
\end{equation}

\noindent The pole orientation is generally unknown, so an observed $\Delta m$ sets an upper limit on $b/a$. 

Figure~\ref{fig:rotprop_vs_size} (right panel) shows maximum lightcurve variability, $\Delta m_\mathrm{max}$, plotted against Centaur size. The larger Centaurs ($D>150$ km) display very ``flat'' lightcurves ($\Delta m\lesssim 0.1$ mag), indicative of more spherical shapes. Bienor is an interesting outlier: with $D\approx 200$ km it has the largest observed variability $\Delta m\sim 0.75$ mag, which implies an axis ratio $\sim$1:2. 

The icy moons of the giant planets larger than $D=400$ km tend to be spherical \citep{2010arXiv1004.1091Lineweaver}, suggesting a transition to a regime where gravitational pull takes over material strength and causes a body to become spherical. The data on Centaurs are sparse, but indicate a transition at smaller sizes, around $D=100$ km. If confirmed, this difference in size at which the gravity becomes dominant should imply different bulk composition and/or internal structures between the two populations.

\subsection{Constraints on density}
\label{sec:densityconstraints}

A useful family of triaxial shapes to consider are the Jacobi ellipsoids. These are the hydrostatic equilibrium shapes of self-gravitating, fluid bodies of uniform density spinning with constant angular frequency \citep{1969efe..book.....Chandra}. The shape (axis ratios) is set by each combination of spin frequency and density, so the latter can be estimated from the lightcurve information, subject to the hydrostatic fluid behaviour assumption.

\begin{figure}
\begin{center}
\includegraphics[width=0.9\textwidth]{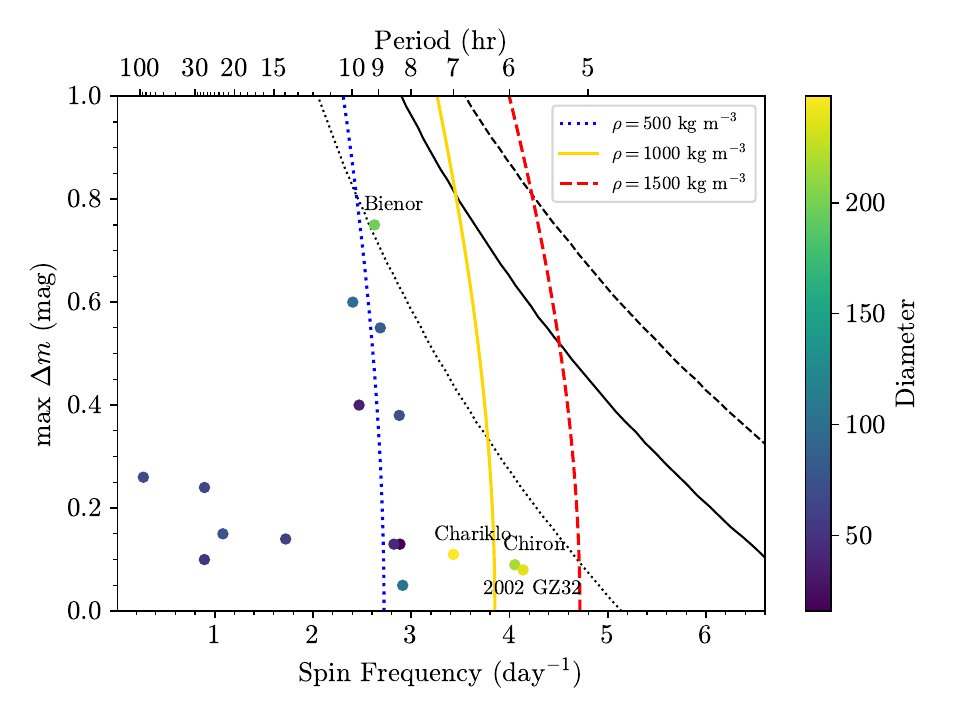}
\end{center}
\caption{Lightcurve variability versus spin rate. Points shaded according to their diameter in km. Three colored lines correspond to Jacobi triaxial ellipsoids with densities 500 (dotted), 1000 (solid) and 1500 kg m$^{-3}$ (dashed), viewed equator-on. Three, more slanted and thinner, black lines indicate the minimum density that balances centripetal and gravitational acceleration at the tips of prolate ellipsoids (dotted, solid and dashed correspond to the same densities as for the colored lines.) \label{fig:spinfreq_vs_dm_vs_rho_vs_size}}
\end{figure}

Figure \ref{fig:spinfreq_vs_dm_vs_rho_vs_size} plots lines of $\Delta m$ against spin frequency for Jacobi ellipsoids of different densities seen equator-on. Centaur lightcurve properties are overplotted and shaded according to each body's size. If Centaurs are in hydrostatic equilibrium, none spins fast enough to require bulk densities much larger than 1000 kg m$^{-3}$. Larger Centaurs ($D \gtrsim 200$ km) concentrate near the density upper limit, mainly due to their fast spins. Smaller Centaurs spin slowly and only require densities of a few hundred kg m$^{-3}$ to sustain their shapes. We note, however, that occultation measurements of Chariklo and its rings have revealed a shape that is not well matched by a figure of hydrostatic equilibrium \citep{2021A&A...652A.141Morgado}. If we simply balance  centrifugal and gravitational accelerations at the tip of prolate ellipsoids to obtain a minimum bulk density, then all Centaurs spin below the critical rate for a bulk density of 500 kg m$^{-3}$ (see black, slanted lines in Fig.\ \ref{fig:spinfreq_vs_dm_vs_rho_vs_size}) which is suggestive of a spin barrier (see section \ref{sec:processes_affecting_rotation}).

\subsection{Binaries, contact binaries and rings}\label{sec:bins}

Two known binaries, (42355) Typhon and Echidna and (65489) Ceto and Phorcys, traverse the Centaur region. They were resolved in HST observations \citep{2006Icar..184..611Noll, 2007Icar..191..286Grundy} with apparent separations 0.1 to 0.2 arcsec ($1330\pm130$ and $1840\pm48$ km, respectively). However, both have orbits extending far beyond Neptune\footnote{$Q_{T\&E}=57.6$ au, $Q_{C\&P}=181.1$ au} and are thus excluded from this review.  None of the Centaurs in Table~\ref{tab:rotational_properties} is a resolved, wide binary \citep[Chapter 9;][]{sickafoosechap9}.

Neither do the Centaur rotational properties suggest strong contact binary candidates, which are usually inferred from large lightcurve variability, $\Delta m>0.9$ mag \citep{2004AJ....127.3023Sheppard,2010ApJ...719.1602Gnat, 2014MNRAS.437.3824Lacerda, 2018AJ....155..248Thirouin}. Bilobed shapes can produce similarly large variation \citep{2016Icar..265...29Descamps}. Contact binaries seem abundant in many small body populations \citep{2004AJ....127.3023Sheppard,2007AJ....134.1133Mann,2011AJ....142...90Lacerda,2018AJ....156..282McNeill,2018AJ....155..248Thirouin,2021Icar..35614098Showalter}. Even though none of the Centaurs in Table~\ref{tab:rotational_properties} displays such extreme variability, Bienor's $\Delta m=0.75$ mag can in principle be caused by an eclipsing binary of two equal spheres. Recent work suggests the contact binary/bilobed shapes observed in JFCs visited by spacecraft can originate in the Centaur region via spin-up due to sublimation torques \citep{2021PSJ.....2...14Safrit}. KBO Arrokoth is evidence that such shape can already exist in the Kuiper Belt. We note that unambiguous shape determination for these objects requires complementary observations. From the ground, dense, multi-chord occultation data may help solve the ambiguity, but only spacecraft flybys can provide ground truth.

A number of factors could contribute to attenuate a lightcurve's range. For instance, the presence of an unknown secondary in unresolved observations from the ground can complicate the interpretation of the lightcurve. The additional cross section, if present, dampens the lightcurve variation giving the impression of a less elongated primary object. Rings such as Chariklo's \citep{2014Natur.508...72BragaRibas}, and possibly Chiron's \citep{2015A&A...576A..18Ortiz,2015Icar..252..271Ruprecht,sickafoosechap9}, and debris or dust in the coma of active Centaurs also contribute additional cross section in unresolved lightcurve observation \citep[Chapter 9;][]{sickafoosechap9} The result is a lower, contaminated lightcurve $\Delta m_c$, given by
\begin{equation}
\label{eq:coma}
    \Delta m_c = -2.5\log\left(\frac{C_\mathrm{min}+C_\mathrm{coma}}{C_\mathrm{max}+C_\mathrm{coma}}\right)=-2.5\log\left(\frac{k+(b/a)}{k+1}\right)
\end{equation}
\noindent where $k=C_\mathrm{coma}/C_\mathrm{max}$ is the rings/coma cross section relative to the maximum nucleus cross section \citep{1990AJ....100..913Luu}. Observations of active Centaurs \citep{1990AJ....100..913Luu,2009AJ....137.4296Jewitt} and Chariklo's ring \citep{2014Natur.508...72BragaRibas} show that $k<0.5$ for the objects in Table~\ref{tab:rotational_properties}. We note that the level of activity (and $k$) varies with time, so they are relevant if occurring at the time the lightcurve is measured. For example, if Chiron's cross-section would include a contemporaneous $k=0.5$ contribution from activity, debris, rings, or a small unresolved companion, then its undiluted lightcurve would vary by $\Delta m=0.14$ mag, compared to the apparent $\Delta m=0.09$ mag. For Bienor, the most photometrically variable Centaur, a relative contribution of $k=0.25$ would imply an undiluted $\Delta m=1.1$ mag. Attempts to detect activity for this Centaur have only rendered an upper limit $k<0.001$ \citep{2009AJ....137.4296Jewitt,2023PSJ.....4...75Dobson}, but rings or an unresolved companion cannot be ruled out. Finally, significant obliquity of the spin pole, combined with an observing geometry away from equinox (equator-on), would also ``hide" an elongated shape (see Eq.~\ref{eq:photometric_range}).

\subsection{Spin vectors and obliquities}


Lightcurve measurements at different aspect angles (the angle between the line of sight and the spin pole) are needed to constrain the spin pole orientation. Such studies of the changing lightcurves of Pholus and Bienor over 13 and 16 years, respectively, have yielded spin poles directed towards ecliptic latitude $\beta = +30\pm 5\deg$ and longitude $\lambda = 145\pm 5 \deg$ for Pholus, and $\beta = 50 \pm 3\deg$ and longitude $\lambda = +35 \pm 8\deg$ for Bienor \citep{2005Icar..175..390Tegler,2017MNRAS.466.4147FernandezValenzuela}. Reliable pole solutions are challenging (reliable lightcurves are needed spanning a wide range of ecliptic longitudes) and hence still rare: Pholus and Bienor are the only cases. For distant objects of moderate inclination, two measurements of $\Delta m$ taken sufficiently far apart along the orbit of an object can set limits on its obliquity, $\varepsilon$, which is the angle between the pole and the normal to the orbit plane \citep{2011AJ....142...90Lacerda}. At $\varepsilon=90\,\deg$ there is a big variation in $\Delta m$ (from a maximum value when seen equator-on down to 0 at pole-on geometry), while for $\varepsilon=0$ there is none.

\begin{figure}
\begin{center}
\includegraphics[width=0.9\textwidth]{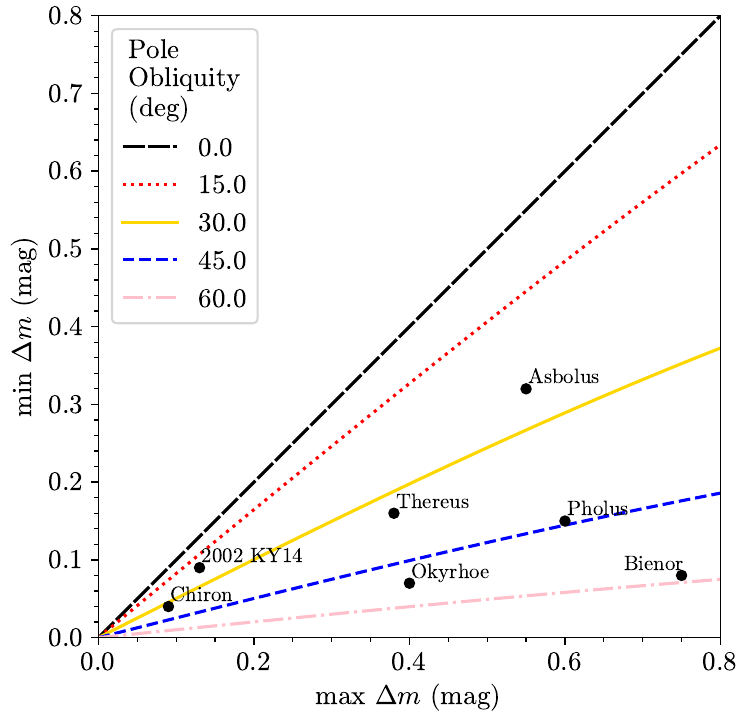}
\end{center}
\caption{Maximum vs minimum lightcurve variability of Centaurs. Lines correspond to Jacobi ellipsoids spinning around the minor principal axis seen equator-on (max $\Delta m$) and at minimum aspect angle along the orbit (min $\Delta m$). \label{fig:obliquity_jacobi_model_no_coma}}
\end{figure}

Figure~\ref{fig:obliquity_jacobi_model_no_coma} plots lines of minimum vs maximum $\Delta m$ for Jacobi ellipsoids of different obliquities. A given shape ($b/a$) implies a maximum $\Delta m=-2.5\log(b/a)$ and the obliquity defines what the minimum $\Delta m$ will be according to Equation~\ref{eq:photometric_range}, where the minimum possible aspect angle is $\theta=90-\varepsilon$ in degrees. Overplotted are the Centaurs with measured minimum and maximum lightcurve ranges, which are respectively upper and lower limits to the real min $\Delta m$ and max $\Delta m$. Under the figure's idealized shape assumptions, the lines indicate minimum obliquity for each Centaur. Pholus's pole obliquity, calculated between the pole solution and the normal to its orbit plane (inclination $i=24.7\deg$ and longitude of ascending node $\Omega=119.4\deg$) is $\varepsilon\approx73\deg$. For Bienor ($i=20.7\deg$ and $\Omega=337.7\deg$) the pole solution implies an obliquity $\varepsilon\approx58\deg$. These do not contradict Fig.~\ref{fig:obliquity_jacobi_model_no_coma}'s minimum obliquities of $\varepsilon_\mathrm{min}\sim45\deg$ for Pholus and $\varepsilon_\mathrm{min}\sim60\deg$ for Bienor. The figure suggests that Centaurs have considerable obliquities. Except Asbolus and 2002 KY14, the remaining 5 Centaurs have obliquities in excess of $30\deg$, and three have obliquities larger than $45\deg$, if their axes ratios are approximately Jacobi-ellipsoidal. Significant obliquity can cause interesting surface illumination and heating patterns, which are important for understanding Centaur activity \citep[see, e.g., Chapters 7 and 8;][]{kokotanekovachap7,bauerchap8}.

\subsection{Processes affecting rotation}
\label{sec:processes_affecting_rotation}
As mentioned above, a number of processes can affect the rotational properties of Centaurs. Collisions, discussed in detail in section \ref{sec:collisions}, will most likely lead to spin-up, which for the smaller objects can lead to rotational breakup \citep{2005PhDT........21Lacerda}. Collisions are unlikely in the Centaur phase, so spins imposed by this process should have been set before the objects left their reservoir in the transneptunian disk.

Torques from cometary activity can cause both spin up or spin down, with equal probability. Such sublimation torques are likely negligible at Centaur distances and certainly for the larger objects, but a scenario in need of mode careful study are those objects that have dipped closer to the sun during their dynamical evolution 
\citep[see, e.g., 39P/Oterma and P/2019 LD2 discussed in Chapter 3;][]{disistochap3}. Sublimative torques may play a role in those cases. 

Processes that lead to the formation of binaries or rings, and their dynamical evolution may also set or affect rotation. This remains largely unexplored for Centaurs, mostly because of the lack of data to constrain any proposed models.

These processes would manifest in the observations in different ways. Collisions cause net spin-up and eventually lead to rotational breakup, setting a spin-barrier and a lower limit to the size of the survivors. Figure~\ref{fig:spinfreq_vs_dm_vs_rho_vs_size} hints at a rotational limit even though the data are still sparse. A simple collisional evolution model in an assumed massive, primordial Kuiper Belt would result in a minimum size of survivors of a few tens of km in radius \citep{2005PhDT........21Lacerda}. Centaurs and descendant JFCs an order of magnitude smaller are observed, but the model is too rudimentary to allow conclusions to be drawn. Sublimative torques lead to a random walk in spin rate, causing the survivor spin rate distribution to be skewed towards slow rotators while also imposing a maximum spin barrier.


\section{Limitations of Measurement Methods}

Many observational techniques to derive physical properties
suffer from limitations. This does not mean that the techniques
are not useful, only that one must interpret results carefully.
We describe some of the particulars here.
Many of these issues are of course common to all small-bodies, not
just to Centaurs. 


As was mentioned earlier, 
visible-wavelength photometry of a bare (inactive)
Centaur can eventually yield an absolute magnitude $H$
but requires an albedo in order to find the diameter.
More specifically, 
the well-known relationship \citep[see][]{2007Icar..190..250P} is 
\begin{equation}
D_{km} = \frac{2a_{km}10^{0.2m_{\odot\lambda}}}{\sqrt{p_\lambda}} 10^{-0.2H_\lambda},
\end{equation}
where $D_{km}$ is the diameter in km, $a_{km}$ is a constant, the number
of km in 1 au, subscript $\lambda$ indicates a quantity at a specific
wavelength, 
$m_{\odot}$ is the monochromatic magnitude of the Sun and 
$p$ is the geometric albedo. Thus we see that
this conversion in fact requires additional information
such as the 
geometric albedo ($p$), and the phase darkening law and
rotational context (for a better $H$).
Extensive photometry on rotational and seasonal timescales can
eventually resolve some of this problem. 

But even overcoming all of these issues,
there is still a problem of what does it
mean to have a diameter of a non-spherical object 
(see Section \ref{sec:shpspin}). With visible
photometry, this is often effectively the geometric-mean of the 
cross-section axes, but this isn't necessarily the best answer.
If the ultimate goal is, say, to understand the mass distribution 
of Centaurs, then one could argue that the `best' diameter 
to use is one that is the geometric mean of the three 
ellipsoid axes, $\sqrt[3]{abc}$.
In a typical scenario, if one measures the
object's light curve, assumes an equator-on
view, and assumes that $b=c$, then one's calculation of the 
volume-derived radius -- i.e., $a(b/a)^{2/3}$ -- is off by
a factor of $\sqrt[3]{c/b}$ from the true answer. 
If every nucleus had the same $c/b$ axial ratio, then this
would just mean that the size distribution would as a whole
be shifted sideways. Of course it is likely the case that
$c/b$ is not constant across objects, and so there could be
a slope bias in the size distribution. This would be especially
deleterious if the largest objects in the size distribution
have $c/b$ values significantly below unity; usually the largest objects in a size distribution are few in number, and if they are widely incorrect, it can significantly affect the resulting power-law exponent. 

\begin{figure}
\begin{center}
\includegraphics[width=0.9\textwidth]{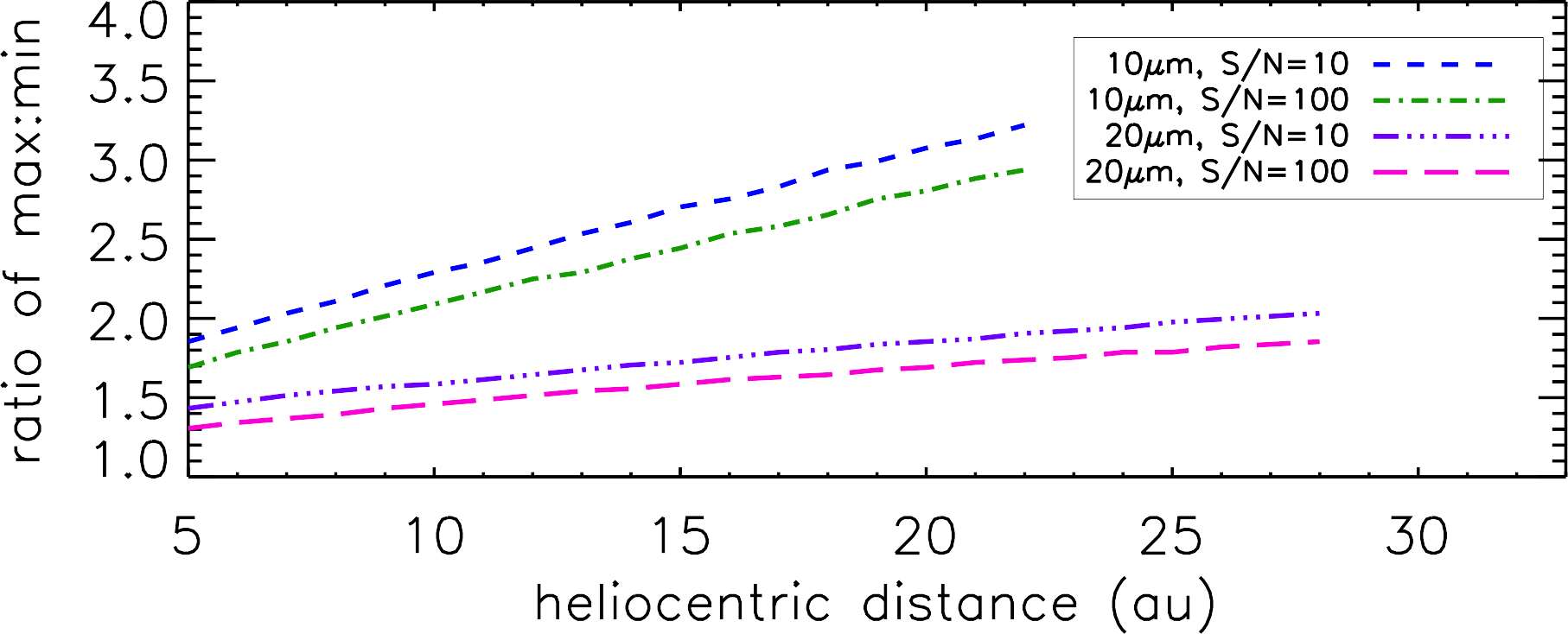}
\end{center}
\caption{Example of the problem with single-band
photometry of the thermal-IR emission from a Centaur.
The plot shows how one-band mid-IR photometry of 
a Centaur at a given heliocentric
distance results in a large uncertainty in the
diameter. The $y$-axis shows the ratio of the
maximum and minimum possible diameters.
In this case, we assumed a geometric albedo of 0.08,
emissivity of $0.95$,
and a phase angle of $2^\circ$. We used NEATM
\citep{1998Icar..131..291H},
and assumed a beaming parameter range of 0.8 to 1.3. 
We tried two wavelengths (10 and 20 $\mu$m) and
two photometric $S/N$ values (100 and 10). 
The 10 $\mu$m lines stop at 22 au since past that,
that wavelength is dominated by reflected sunlight. 
The 20 $\mu$m lines stop at 28 au since a phase
angle of 2 degrees is impossible past that. 
For more distant (and thus cooler) Centaurs, the
range of possible diameters spans a large 
factor. This uncertainty ignores the applicability of NEATM
in the first place, since more distant objects are
more likely to be fast rotators.  \label{fig:wien}}
\end{figure}

To return to the problem of determining the size, certainly
infrared techniques, using measurements of the thermal radiation,
have given us great improvements. Since the square of the diameter is,
broadly speaking, proportional to the thermal infrared
flux density divided by $(1-pq_{ph})$, where $q_{ph}$ is the phase
integral, and since $pq_{ph}$ is generally small, the systematic
uncertainty is in principle lower. Nonetheless, the
rotational context can still be a problem, and the phase
darkening of thermal emission is not well understood. 
The temperature map of a Centaur's surface may not be
well known, either because the object's shape is poorly
constrained or because the fundamental thermal 
quantities (thermal inertia, thermal diffusivity) are
not well constrained, or both. 
Detailed work on the thermal properties of near-Earth
asteroids \citep[e.g.,][]{2007astro.ph..3085W, 2018arXiv181101454W, 2018Icar..303..220H, 2018AJ....155...74M, 2023PSJ.....4....5M} reveals the simplistic
thermal models to interpret single-snapshot thermal-IR photometry
of highly nonspherical objects can
give diameters that are significantly
off. Such a problem can be even more severe for
distant objects like Centaurs where photometry in the
vicinity of 10 $\mu$m lies on the Wien-law side of a
thermal continuum. This is demonstrated for a particular
case in Figure~\ref{fig:wien}, where there is just
one band of mid-IR photometry and so NEATM
and a beaming parameter range is assumed. 
For example, suppose we have thermal-IR photometry at a wavelength
of 10 microns that has yielded a $S/N=10$ measurement of
an object that is 10 au from the Sun. Assuming 
that the beaming parameter is between 0.8 and 1.3,
we find from the plot
that a wide range of radii fit the data point, with the maximum
possible radius being ${\cal R}=2.3$ times larger than the minimum. 
This corresponds to a fractional
uncertainty in the radius of $({\cal R}-1)/({\cal R}+1) = 39$\%, 
quite significant.
The figure shows that the result has some dependence on
$S/N$, since less accurate photometry will let more 
possible values of NEATM parameters work, but this problem
persists even for high quality photometry ($S/N=100$ in the plot).
In any case, this demonstrates that multiband 
photometry improves the ability to apply more
sophisticated thermal models.


As discussed in Eq.~\ref{eq:coma}, a Centaur's coma will 
complicate finding a Centaur's physical properties. 
In principle
an extraction technique can be done 
\citep{1995A&A...293L..43L,1999Icar..140..189L,2018PASP..130j4501H}
but the
method loses reliability the coarser the spatial
resolution of the imaging and the more dominant the
coma's flux is against the nucleus's. 
While traditionally one of the criteria for
determining if an object has cometary activity is
to look for extended emission beyond the PSF
\citep[(see, e.g.,][]{1991ASSL..167...19J}, it is
possible for comets to be active and yet hide
coma within the PSF, as is the case apparently
for ecliptic-comet 2P/Encke \citep{2005Icar..175..194F}. 
For more distant objects such as 
Centaurs, it would be even easier to be fooled
into thinking that an ostensibly PSF-looking 
Centaur is bare when it in fact has some cometary
activity. 

We mention one further limitation here: the spatial resolution
limitation for identifying a Centaur as a binary 
(cf. \S\ref{sec:bins}). HST has been used to show
that at least about one-fifth of cold classical TNOs are binaries
\citep{2008Icar..194..758N,2024PSJ.....5..143P}, 
with the limitation there for detection
being that the secondary needs to be
bright enough and at least several hundred kilometers
from the primary. Searches that use HST to 
look for binarity among Centaurs
include, e.g., that of \cite{2020AJ....159..209L}, who 
looked at 23 objects that would be included in our Centaur
definition here (among many other objects)
but found no binaries, even though
their linear spatial resolution was roughly about
twice as good as it is for cold-classical TNOs. 
In any case, if there is a population of `tight'
Centaur binaries, it may be necessary
to conduct a search where spatial resolutions better than
$\sim$100 km are possible.
This is 14 milliarcsec for an object 10 au away, i.e. 
just under 0.5 JWST NIRCam short-wavelength pixels.
It would certainly be interesting to determine definitively
whether the binary fraction
of the Centaurs were much lower than that of the TNO population.


\section{Comparative Planetology of Outer
Solar System Bodies}

The term ``comparative planetology'' evokes an approach of taking a set of diverse measurements and properties of individual objects in search of a deeper understanding of the set.  Much of what we call comparative planetology has been driven by comparing and contrasting terrestrial planets, giant planets, dwarf planets where individual properties matter a great deal.  When applying this notion to small bodies, a different framework and approach becomes important.

For small bodies, individual properties are still important but the distribution of those properties across the full population provide far more insight.  A fantastic example of this approach is the data from the Sloan Digital Sky Survey (SDSS) \citep{2001AJ....122.2749I,2002AJ....124.2943I}.  A prominent example of this work is showing the compositional ties within and between asteroid collisional families.
While we do not yet have a dataset comparable to SDSS that covers the entire Solar System, we do expect such fundamental data to come from future surveys.  In the meantime, we make do with such data as we have.  The critical point here is to focus on the properties of populations of small bodies and look for insight in comparing and contrasting populations.  For this section, comparative planetology will refer to the comparison of population properties and what it can tell us.

\subsection{Centaur Shapes}

The concept of a ``primordial'' object is unfortunately an overused and poorly codified term used widely in planetary science.  At one time or another the term primordial has probably been applied to every small body in the Solar System.

With the successful flyby of Arrokoth by New Horizons, we finally have an object truly worthy of the term.  As discussed by \cite{2020Sci...367.6620M}, Arrokoth is now helping to more clearly define the concept of what a primordial object is.  Here we have an example of an object that has remained essentially untouched since the end of the accretion phase of building our Solar System (or at least since the time of thermal equilibration once the disk cleared).  As such, it directly informs and constrains theories of how the accretion process works.  As a cold-classical Kuiper Belt object (CCKBO), we can look upon these objects as being truly primordial.  It would appear that every other population of small body, while made up of primordial material, has been more affected by post-accretional processes than what we understand for CCKBOs.

An important question to address is the extent and nature of any processing that has been at work on Centaurs.  Quantifying the amount of processing will go far in placing them in context within the other populations of small bodies and help identify the source population and pathway by which an object becomes a Centaur.

Our best prospect, short of spacecraft missions, for assessing the primordial nature of Centaurs lies with detailed stellar occultation observations.  Observations of Jupiter Trojans conducted in support of the Lucy mission show that all the targets have significantly more complicated shapes than exhibited by Arrokoth
\citep[e.g.,][]{2021PSJ.....2..202B,2023PSJ.....4...18M,2024SSRv..220...17M}.  In this case, this constraint is not about the contact binary shape of Arrokoth.  None of the Lucy targets is an obvious contact binary though we cannot exclude some being badly distorted contact binaries. The more interesting constraint is that the Lucy targets do not have smooth (i.e., elliptical) shapes and all show significant modification with respect to Arrokoth.  This result runs over the range of 30-120 km for the Lucy targets.  It remains to be seen if there are any remaining signatures of the primordial shape for these objects and the definitive answers will come with the culmination of that mission.

There are three key observables in the shape of a small body that are likely to be important clues tracing back to primordial shape and thus the degree of processing that Centaurs have undergone:  (1) a contact binary shape that is consistent with a gentle merger
\citep[though see][]{2021PSJ.....2...14Safrit}, (2) highly oblate shape, and (3) a smooth shape.  There may be other signatures we do not yet know enough about, such as the frequency of tight binaries, since we do not have good knowledge of this for any other distant population.

\subsection{Collisions}
\label{sec:collisions}
Collisions can significantly alter small body populations.
This is particularly evident in the main asteroid belt.
Its size distribution shows clear signs of being sculpted by collisions \citep{Bottke2005Icar, Bottke2020AJ}.
Furthermore, the main asteroid belt contains many collisional families, which can be traced back to catastrophic disruptions of larger bodies \citep[e.g.,][]{Vokrouhlicky2010AJ, Licandro2012A&A, Spoto2015Icar, Bolin2017Icar}.
In contrast, the Jupiter Trojans have only one major collisional family, the one of Eurybates \citep{BrozRozehnal2011, Marschall2022AJ}.

The amount of collisional evolution is driven by the dynamical evolution of a small body population.
There is now a general consensus that planetesimals that end up as Centaurs and JFCs formed in an original reservoir, the primordial Kuiper belt (PKB), located between 20-40 au from the Sun \citep[e.g.,][]{Nesvorny2018ARA&A}. 
From there, they were scattered into the current trans-Neptunian region. 
From that region, the scattered disk in particular acts as the source reservoir for Centaurs and JFCs \citep[e.g.,][]{Duncan1997Sci}. 
Crucially, the scattering phase of planetesimals must occur as the gas disk dissipates to prevent the damping of the scattered orbits. 
This is well described by the final phase of planetesimal-driven migration, particularly Neptune's migration \citep[e.g.,][]{Nesvorny2018NatAs,Nesvorny2021ApJL}. 

The collisional environment of a Centaur can thus be divided into three phases: 1) the period in the PKB and subsequent scattering into the current trans-Neptunian region; 2) the period in the scattered disk; 3) the Centaur phase \citep{Bottke2023PSJ}.
Phase 1 in the PKB may only last a few tens of millions of years.
Nevertheless, it is the phase with the largest number of impacts because of the large mass of the PKB (several tens of Earth masses) compared to the scattered disk \citep[1/400 of the initial PKB;][]{Bottke2020AJ, Nesvorny2020AJ}.
Phase 2 lasts for most of the lifetime of the Solar System (4.5 Gyr) and may still contribute moderately to the collisional evolution.
The final phase, when objects are scattered into the giant planet region, lasts only a few million years and will not contribute in any notable way to the collisional evolution of Centaurs.
In this phase, sublimation-driven activity will dominate any collisional evolution 
\citep[see, e.g., Chapters 5, 7, and 8;][]{peixinhochap5,kokotanekovachap7,bauerchap8}.

Another crucial ingredient to understanding collisional evolution is the initial size distribution.
Current planetesimal formation models suggest that planetesimals form via the streaming instability.
This disk instability accumulates dust particles in clouds massive enough for self-gravity to cause the cloud to collapse and directly form large planetesimals. 
This mechanism predicts an initial Gaussian size distribution centred around D = 100 km \citep[e.g.,][]{Simon2016ApJ,Schafer2017A&A,Klahr2020ApJ}.
While these are models, the current Kuiper belt has a distinct bump around 100 km.
This is in line with a leftover streaming instability size distribution.

\begin{figure}
    \centering
    \includegraphics[width=0.9\linewidth]{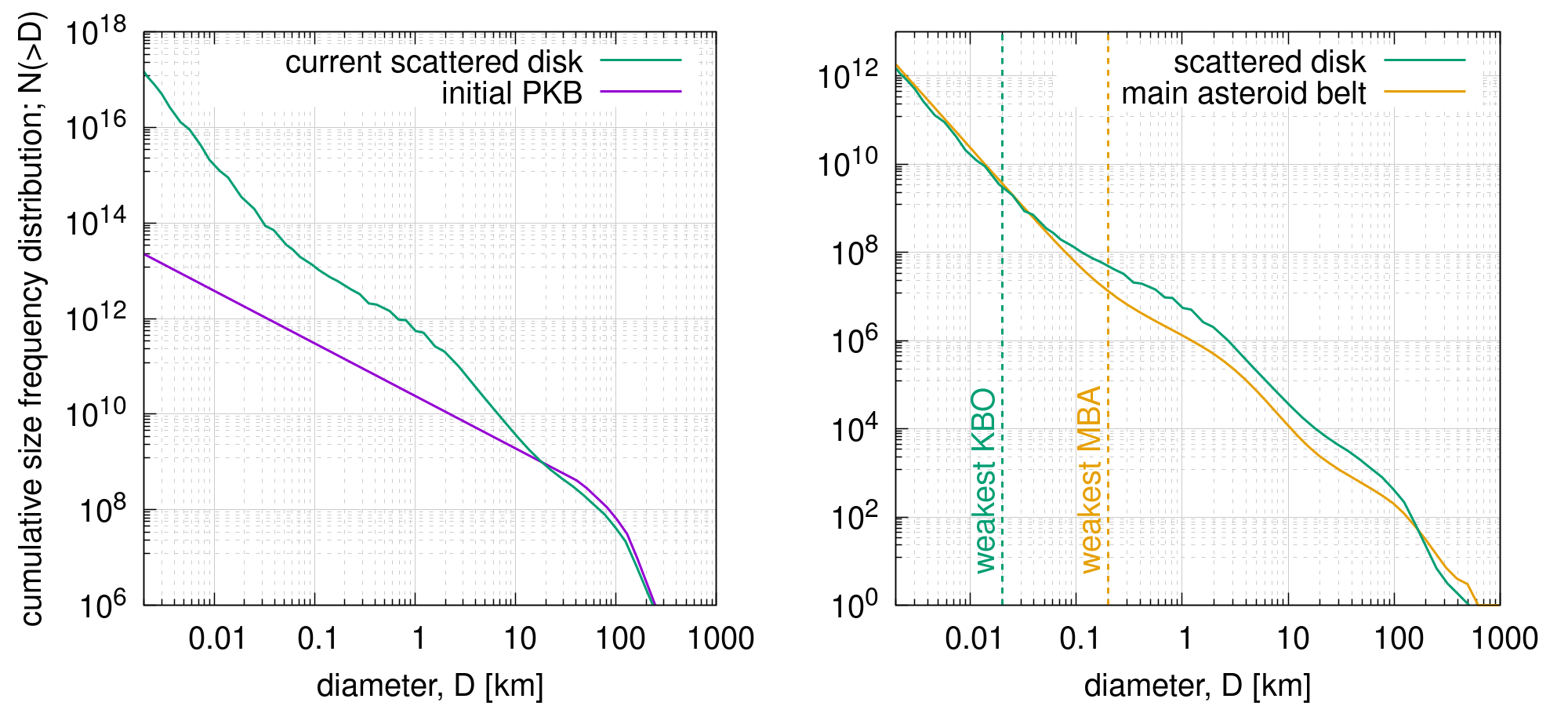}
    \caption{The left panel shows the cumulative size frequency distribution (CSFD) of the initial primordial Kuiper-belt (PKB) and the collisionally evolved CSFD of the current day scattered disk according to \cite{Bottke2023PSJ}. The dynamical depletion of the current scattered disk is not shown.
    The right panel shows the CSFD of the current day scattered disk \citep{Bottke2023PSJ} and the main asteroid belt \citep{Bottke2020AJ}. The dashed lines show the approximate location of the weakest Kuiper-belt object (KBO) and main belt asteroid (MBA). The scattered disk SFD has been scaled to be on the same order of magnitude of the main belt.}
    \label{fig:KBO-SFD}
\end{figure}

\cite{Bottke2023PSJ} showed that an initial streaming instability size distribution can indeed evolve into the current scattered disk size distribution.
They showed that this evolution is consistent with (i) crater SFDs on icy satellites and KBOs \citep[e.g.,][]{Singer2019Sci, Zahnle2003Icar, Schenk2004jpsm, Kirchoff2010Icar}, and (ii) observed SFDs of populations derived from the PKB (e.g., Jupiter’s Trojans) \citep{Wong2015AJ,Yoshida2017AJ}. 
The craters on icy satellites were used to infer the size distribution of the PKB population scattered onto planet-crossing orbits and those that went to the scattered disk.
Figure~\ref{fig:KBO-SFD} shows the initial and evolved Kuiper belt size distribution.
It exhibits a wavy shape characteristic of collisional evolution.

The Kuiper belt thus appears to be similarly collisionally evolved as the main asteroid belt with one significant difference (Fig.~\ref{fig:KBO-SFD}).
While the weakest body in the asteroid belt has a diameter of roughly 200 m, it is only 20 m in the scattered disk (Fig.~\ref{fig:KBO-SFD}).
This shift in the weakest bodies shifts the location of the ``collisional wave" of the size distribution of asteroids to KBOs to smaller sizes.

The fact that the scattered disk size distribution can be explained by collisionally evolving a streaming instability distribution indicates that most smaller bodies in the scattered disk, and by extension the Centaur population, are collisional fragments.
\cite{Marschall2023ACM} argue that 50\% of scattered disk objects/comets/centaurs larger than 10 km should be collisional fragments. 
For objects larger than 1 km, at least 95\% of objects are fragments.
In contrast, large Centaurs (100 km and larger) are, to a large extent, untouched by disruption and thus the most ``pristine" in this respect.


\section{Future Prospects}

As with many things in astronomy, one way to address some of the
issues we have described above is with large amounts of data, and
especially data that sample
new regimes of faintness. Number statistics are crucial. 
New datasets from Vera C. Rubin 
Observatory and NEO Surveyor promise to improve
the discovery rate and characterization of many Centaurs
\citep{2023PSJ.....4..224M,2023ApJS..266...22S,efvchap14}. 
Further data from existing surveys like Pan-STARRS will
continue to help us with photometric coverage. This is
especially useful for objects like Centaurs that tend
to have multidecade orbital periods, so it takes a while
for our vantage point (i.e. sub-Earth latitude) to change
significantly. Disentangling the photometric
manifestations of a Centaur's shape, rotation
period, spin axis direction, color/albedo variations, and 
phase darkening can be challenging, just as it is
with any asteroid. Add onto that the possibility that
some Centaurs could have hiccups or outbursts of
activity, that may or not be resolved, and it is clear
that denser photometric coverage, at a range
of cadences, can be necessary.

Insights into cometary activity \citep[see Chapter 8;][]{bauerchap8}, 
especially such
activity at distances from the Sun where water ice sublimation
is {\sl not} the main driver, will be useful for
further studies of how Centaur surfaces change. This can
include more detailed assessment of Rosetta's observations
of 67P when that comet was far from the Sun. But it may
also require sending a spacecraft to study a Centaur
up close \citep[see Chapter 16;][]{harrissternchap16}.
Once we can treat a Centaur as a geologic object it will surely
improve our understanding of the observations of a Centaur
as an astronomical object.

\begin{acknowledgments}
We thank two
anonymous referees whose helpful 
reviews improved our manuscript.
\end{acknowledgments}

\bibliography{nucleuschapter-for-arxiv}{}

\begin{thebibliography}{}
\expandafter\ifx\csname natexlab\endcsname\relax\def\natexlab#1{#1}\fi
\providecommand{\url}[1]{\href{#1}{#1}}
\providecommand{\dodoi}[1]{doi:~\href{http://doi.org/#1}{\nolinkurl{#1}}}
\providecommand{\doeprint}[1]{\href{http://ascl.net/#1}{\nolinkurl{http://ascl.net/#1}}}
\providecommand{\doarXiv}[1]{\href{https://arxiv.org/abs/#1}{\nolinkurl{https://arxiv.org/abs/#1}}}

\bibitem[{J. Bauer {et~al.}(2025)Bauer, Ivanova, McKay, \& Sarid}]{bauerchap8}
Bauer, J., Ivanova, O., McKay, A., \& Sarid, G. 2025, \bibinfo{title}{{Activity, Outbursts and Explosions},} in Centaurs, ed. K.~{Volk}, M.~{Womack}, \& J.~{Steckloff} (IOP Publishing), 8--1\textemdash8--23, \dodoi{10.1088/2514-3433/ada267ch8}

\bibitem[{B.~T. {Bolin} {et~al.}(2017){Bolin}, {Delbo}, {Morbidelli}, \& {Walsh}}]{Bolin2017Icar}
{Bolin}, B.~T., {Delbo}, M., {Morbidelli}, A., \& {Walsh}, K.~J. 2017, \bibinfo{title}{{Yarkovsky V-shape identification of asteroid families},} \icarus, 282, 290, \dodoi{10.1016/j.icarus.2016.09.029}

\bibitem[{W.~F. {Bottke} {et~al.}(2005){Bottke}, {Durda}, {Nesvorn{\'y}}, {Jedicke}, {Morbidelli}, {Vokrouhlick{\'y}}, \& {Levison}}]{Bottke2005Icar}
{Bottke}, W.~F., {Durda}, D.~D., {Nesvorn{\'y}}, D., {et~al.} 2005, \bibinfo{title}{{The fossilized size distribution of the main asteroid belt},} \icarus, 175, 111, \dodoi{10.1016/j.icarus.2004.10.026}

\bibitem[{W.~F. {Bottke} {et~al.}(2020){Bottke}, {Vokrouhlick{\'y}}, {Ballouz}, {Barnouin}, {Connolly}, {Elder}, {Marchi}, {McCoy}, {Michel}, {Nolan}, {Rizk}, {Scheeres}, {Schwartz}, {Walsh}, \& {Lauretta}}]{Bottke2020AJ}
{Bottke}, W.~F., {Vokrouhlick{\'y}}, D., {Ballouz}, R.~L., {et~al.} 2020, \bibinfo{title}{{Interpreting the Cratering Histories of Bennu, Ryugu, and Other Spacecraft-explored Asteroids},} \aj, 160, 14, \dodoi{10.3847/1538-3881/ab88d3}

\bibitem[{W.~F. {Bottke} {et~al.}(2023){Bottke}, {Vokrouhlick{\'y}}, {Marshall}, {Nesvorn{\'y}}, {Morbidelli}, {Deienno}, {Marchi}, {Dones}, \& {Levison}}]{Bottke2023PSJ}
{Bottke}, W.~F., {Vokrouhlick{\'y}}, D., {Marshall}, R., {et~al.} 2023, \bibinfo{title}{{The Collisional Evolution of the Primordial Kuiper Belt, Its Destabilized Population, and the Trojan Asteroids},} \psj, 4, 168, \dodoi{10.3847/PSJ/ace7cd}

\bibitem[{F. {Braga-Ribas} {et~al.}(2014){Braga-Ribas}, {Sicardy}, {Ortiz}, {Snodgrass}, {Roques}, {Vieira-Martins}, {Camargo}, {Assafin}, {Duffard}, {Jehin}, {Pollock}, {Leiva}, {Emilio}, {Machado}, {Colazo}, {Lellouch}, {Skottfelt}, {Gillon}, {Ligier}, {Maquet}, {Benedetti-Rossi}, {Gomes}, {Kervella}, {Monteiro}, {Sfair}, {El Moutamid}, {Tancredi}, {Spagnotto}, {Maury}, {Morales}, {Gil-Hutton}, {Roland}, {Ceretta}, {Gu}, {Wang}, {Harps{\o}e}, {Rabus}, {Manfroid}, {Opitom}, {Vanzi}, {Mehret}, {Lorenzini}, {Schneiter}, {Melia}, {Lecacheux}, {Colas}, {Vachier}, {Widemann}, {Almenares}, {Sandness}, {Char}, {Perez}, {Lemos}, {Martinez}, {J{\o}rgensen}, {Dominik}, {Roig}, {Reichart}, {Lacluyze}, {Haislip}, {Ivarsen}, {Moore}, {Frank}, \& {Lambas}}]{2014Natur.508...72BragaRibas}
{Braga-Ribas}, F., {Sicardy}, B., {Ortiz}, J.~L., {et~al.} 2014, \bibinfo{title}{{A ring system detected around the Centaur (10199) Chariklo},} \nat, 508, 72, \dodoi{10.1038/nature13155}

\bibitem[{M. {Bro{\v{z}}} \& J. {Rozehnal}(2011){Bro{\v{z}}} \& {Rozehnal}}]{BrozRozehnal2011}
{Bro{\v{z}}}, M., \& {Rozehnal}, J. 2011, \bibinfo{title}{{Eurybates - the only asteroid family among Trojans?},} \mnras, 414, 565, \dodoi{10.1111/j.1365-2966.2011.18420.x}

\bibitem[{M.~W. {Buie} {et~al.}(2018){Buie}, {Zangari}, {Marchi}, {Levison}, \& {Mottola}}]{2018AJ....155..245B}
{Buie}, M.~W., {Zangari}, A.~M., {Marchi}, S., {Levison}, H.~F., \& {Mottola}, S. 2018, \bibinfo{title}{{Light Curves of Lucy Targets: Leucus and Polymele},} \aj, 155, 245, \dodoi{10.3847/1538-3881/aabd81}

\bibitem[{M.~W. {Buie} {et~al.}(2021){Buie}, {Keeney}, {Strauss}, {Blank}, {Moore}, {Porter}, {Wasserman}, {Weryk}, {Levison}, {Olkin}, {Leiva}, {Bardecker}, {Brown}, {Brown}, {Collins}, {Davidson}, {Dunham}, {Dunham}, {Eaccarino}, {Finley}, {Fuller}, {Garcia}, {George}, {Getrost}, {Gialluca}, {Givot}, {Gupton}, {Hanna}, {Hergenrother}, {Hernandez}, {Hill}, {Hinton}, {Holt}, {Howell}, {Jewell}, {Kamin}, {Kammer}, {Kareta}, {Kayl}, {Keller}, {Kenyon}, {Kester}, {Kidd}, {Lauer}, {Leung}, {Lorusso}, {Lundgren}, {Magana}, {Maley}, {Marchis}, {Marcialis}, {McCandless}, {McCrystal}, {McGraw}, {Miller}, {Mueller}, {Noonan}, {Olsen}, {Patton}, {Peluso}, {Person}, {Rigby}, {Rolfsmeier}, {Salmon}, {Samaniego}, {Sawyer}, {Schulz}, {Skrutskie}, {Smith}, {Spencer}, {Springmann}, {Stanbridge}, {Stoffel}, {Tamblyn}, {Tobias}, {Verbiscer}, {von Schalscha}, {Werts}, \& {Zhang}}]{2021PSJ.....2..202B}
{Buie}, M.~W., {Keeney}, B.~A., {Strauss}, R.~H., {et~al.} 2021, \bibinfo{title}{{Size and Shape of (11351) Leucus from Five Occultations},} \psj, 2, 202, \dodoi{10.3847/PSJ/ac1f9b}

\bibitem[{S. {Chandrasekhar}(1969){Chandrasekhar}}]{1969efe..book.....Chandra}
{Chandrasekhar}, S. 1969, {Ellipsoidal figures of equilibrium} (Yale University Press, New Haven, CT)

\bibitem[{A. {Delsanti} {et~al.}(2004){Delsanti}, {Hainaut}, {Jourdeuil}, {Meech}, {Boehnhardt}, \& {Barrera}}]{2004A&A...417.1145Delsanti}
{Delsanti}, A., {Hainaut}, O., {Jourdeuil}, E., {et~al.} 2004, \bibinfo{title}{{Simultaneous visible-near IR photometric study of Kuiper Belt Object surfaces with the ESO/Very Large Telescopes},} \aap, 417, 1145, \dodoi{10.1051/0004-6361:20034182}

\bibitem[{P. {Descamps}(2016){Descamps}}]{2016Icar..265...29Descamps}
{Descamps}, P. 2016, \bibinfo{title}{{Near-equilibrium dumb-bell-shaped figures for cohesionless small bodies},} \icarus, 265, 29, \dodoi{10.1016/j.icarus.2015.10.010}

\bibitem[{R.~P. {Di Sisto} {et~al.}(2025){Di Sisto}, Gallardo, \& Dones}]{disistochap3}
{Di Sisto}, R.~P., Gallardo, T., \& Dones, L. 2025, \bibinfo{title}{{Dynamics of Centaurs and Links to Scattering and Comet Populations},} in Centaurs, ed. K.~{Volk}, M.~{Womack}, \& J.~{Steckloff} (IOP Publishing), 3--1\textemdash3--28, \dodoi{10.1088/2514-3433/ada267ch3}

\bibitem[{M.~M. {Dobson} {et~al.}(2023){Dobson}, {Schwamb}, {Benecchi}, {Verbiscer}, {Fitzsimmons}, {Shingles}, {Denneau}, {Heinze}, {Smith}, {Tonry}, {Weiland}, \& {Young}}]{2023PSJ.....4...75Dobson}
{Dobson}, M.~M., {Schwamb}, M.~E., {Benecchi}, S.~D., {et~al.} 2023, \bibinfo{title}{{Phase Curves of Kuiper Belt Objects, Centaurs, and Jupiter-family Comets from the ATLAS Survey},} \psj, 4, 75, \dodoi{10.3847/PSJ/acc463}

\bibitem[{R. {Duffard} {et~al.}(2014){Duffard}, {Pinilla-Alonso}, {Santos-Sanz}, {Vilenius}, {Ortiz}, {Mueller}, {Fornasier}, {Lellouch}, {Mommert}, {Pal}, {Kiss}, {Mueller}, {Stansberry}, {Delsanti}, {Peixinho}, \& {Trilling}}]{2014AandA...564A..92Duffard}
{Duffard}, R., {Pinilla-Alonso}, N., {Santos-Sanz}, P., {et~al.} 2014, \bibinfo{title}{{``TNOs are Cool'': A survey of the trans-Neptunian region. XI. A Herschel-PACS view of 16 Centaurs},} \aap, 564, A92, \dodoi{10.1051/0004-6361/201322377}

\bibitem[{M. {Duncan} {et~al.}(2004){Duncan}, {Levison}, \& {Dones}}]{2004come.book..193D}
{Duncan}, M., {Levison}, H., \& {Dones}, L. 2004, \bibinfo{title}{{Dynamical evolution of ecliptic comets},} in Comets II, ed. M.~C. {Festou}, H.~U. {Keller}, \& H.~A. {Weaver} (Univ. Arizona Press, Tucson, AZ), 193

\bibitem[{M.~J. {Duncan} \& H.~F. {Levison}(1997){Duncan} \& {Levison}}]{Duncan1997Sci}
{Duncan}, M.~J., \& {Levison}, H.~F. 1997, \bibinfo{title}{{A scattered comet disk and the origin of Jupiter family comets},} Science, 276, 1670, \dodoi{10.1126/science.276.5319.1670}

\bibitem[{Y.~R. {Fern{\'a}ndez} {et~al.}(2005){Fern{\'a}ndez}, {Lowry}, {Weissman}, {Mueller}, {Samarasinha}, {Belton}, \& {Meech}}]{2005Icar..175..194F}
{Fern{\'a}ndez}, Y.~R., {Lowry}, S.~C., {Weissman}, P.~R., {et~al.} 2005, \bibinfo{title}{{New near-aphelion light curves of Comet 2P/Encke},} \icarus, 175, 194, \dodoi{10.1016/j.icarus.2004.10.019}

\bibitem[{E. {Fern\'andez-Valenzuela} {et~al.}(2025){Fern\'andez-Valenzuela}, {Guilbert-Lepoutre}, Schwamb, Holler, Rommel, \& Schambeau}]{efvchap14}
{Fern\'andez-Valenzuela}, E., {Guilbert-Lepoutre}, A., Schwamb, M.~E., {et~al.} 2025, \bibinfo{title}{{New Observational Studies: Early Results from JWST and Future Prospects with JWST and the Vera C. Rubin Observatory},} in Centaurs, ed. K.~{Volk}, M.~{Womack}, \& J.~{Steckloff} (IOP Publishing), 14--1\textemdash14--16, \dodoi{10.1088/2514-3433/ada267ch14}

\bibitem[{E. {Fern{\'a}ndez-Valenzuela} {et~al.}(2017){Fern{\'a}ndez-Valenzuela}, {Ortiz}, {Duffard}, {Morales}, \& {Santos-Sanz}}]{2017MNRAS.466.4147FernandezValenzuela}
{Fern{\'a}ndez-Valenzuela}, E., {Ortiz}, J.~L., {Duffard}, R., {Morales}, N., \& {Santos-Sanz}, P. 2017, \bibinfo{title}{{Physical properties of centaur (54598) Bienor from photometry},} \mnras, 466, 4147, \dodoi{10.1093/mnras/stw3264}

\bibitem[{S. {Fornasier} {et~al.}(2013){Fornasier}, {Lellouch}, {M{\"u}ller}, {Santos-Sanz}, {Panuzzo}, {Kiss}, {Lim}, {Mommert}, {Bockel{\'e}e-Morvan}, {Vilenius}, {Stansberry}, {Tozzi}, {Mottola}, {Delsanti}, {Crovisier}, {Duffard}, {Henry}, {Lacerda}, {Barucci}, \& {Gicquel}}]{2013AandA...555A..15Fornasier}
{Fornasier}, S., {Lellouch}, E., {M{\"u}ller}, T., {et~al.} 2013, \bibinfo{title}{{TNOs are Cool: A survey of the trans-Neptunian region. VIII. Combined Herschel PACS and SPIRE observations of nine bright targets at 70-500 {\ensuremath{\mu}}m},} \aap, 555, A15, \dodoi{10.1051/0004-6361/201321329}

\bibitem[{W.~C. {Fraser} {et~al.}(2024){Fraser}, {Dones}, {Volk}, {Womack}, \& {Nesvor{\'y}}}]{2024come.book..121F}
{Fraser}, W.~C., {Dones}, L., {Volk}, K., {Womack}, M., \& {Nesvor{\'y}}, D. 2024, \bibinfo{title}{{The Transition from the Kuiper Belt to the Jupiter-Family Comets},} in Comets III, ed. K.~J. {Meech}, M.~R. {Combi}, D.~{Bockel{\'e}e-Morvan}, S.~N. {Raymodn}, \& M.~E. {Zolensky} (Univ. Arizona Press, Tucson, AZ), 121--152

\bibitem[{O. {Gnat} \& R. {Sari}(2010){Gnat} \& {Sari}}]{2010ApJ...719.1602Gnat}
{Gnat}, O., \& {Sari}, R. 2010, \bibinfo{title}{{Equilibrium Configurations of Synchronous Binaries: Numerical Solutions and Application to Kuiper Belt Binary 2001 QG$_{298}$},} \apj, 719, 1602, \dodoi{10.1088/0004-637X/719/2/1602}

\bibitem[{W.~M. {Grundy} {et~al.}(2007){Grundy}, {Stansberry}, {Noll}, {Stephens}, {Trilling}, {Kern}, {Spencer}, {Cruikshank}, \& {Levison}}]{2007Icar..191..286Grundy}
{Grundy}, W.~M., {Stansberry}, J.~A., {Noll}, K.~S., {et~al.} 2007, \bibinfo{title}{{The orbit, mass, size, albedo, and density of (65489) Ceto/Phorcys: A tidally-evolved binary Centaur},} \icarus, 191, 286, \dodoi{10.1016/j.icarus.2007.04.004}

\bibitem[{A.~W. {Harris}(1998){Harris}}]{1998Icar..131..291H}
{Harris}, A.~W. 1998, \bibinfo{title}{{A Thermal Model for Near-Earth Asteroids},} \icarus, 131, 291, \dodoi{10.1006/icar.1997.5865}

\bibitem[{W. Harris {et~al.}(2025)Harris, Stern, \& Villanueva}]{harrissternchap16}
Harris, W., Stern, S.~A., \& Villanueva, G.~L. 2025, \bibinfo{title}{{Centaur Missions},} in Centaurs, ed. K.~{Volk}, M.~{Womack}, \& J.~{Steckloff} (IOP Publishing), 16--1\textemdash16--26, \dodoi{10.1088/2514-3433/ada267ch16}

\bibitem[{M. Hirabayashi {et~al.}(2025)Hirabayashi, S\'anchez, \& Sarid}]{hirabayashichap12}
Hirabayashi, M., S\'anchez, P., \& Sarid, G. 2025, \bibinfo{title}{{Centaurs’ Deep Interiors},} in Centaurs, ed. K.~{Volk}, M.~{Womack}, \& J.~{Steckloff} (IOP Publishing), 12--1\textemdash12--30, \dodoi{10.1088/2514-3433/ada267ch12}

\bibitem[{E.~S. {Howell} {et~al.}(2018){Howell}, {Magri}, {Vervack}, {Nolan}, {Taylor}, {Fern{\'a}ndez}, {Hicks}, {Somers}, {Lawrence}, {Rivkin}, {Marshall}, \& {Crowell}}]{2018Icar..303..220H}
{Howell}, E.~S., {Magri}, C., {Vervack}, R.~J., {et~al.} 2018, \bibinfo{title}{{SHERMAN - A shape-based thermophysical model II. Application to 8567 (1996 HW1)},} \icarus, 303, 220, \dodoi{10.1016/j.icarus.2017.12.003}

\bibitem[{M.-T. {Hui} \& J.-Y. {Li}(2018){Hui} \& {Li}}]{2018PASP..130j4501H}
{Hui}, M.-T., \& {Li}, J.-Y. 2018, \bibinfo{title}{{Is the Cometary Nucleus-extraction Technique Reliable?},} \pasp, 130, 104501, \dodoi{10.1088/1538-3873/aad538}

\bibitem[{{\v{Z}}. {Ivezi{\'c}} {et~al.}(2001){Ivezi{\'c}}, {Tabachnik}, {Rafikov}, {Lupton}, {Quinn}, {Hammergren}, {Eyer}, {Chu}, {Armstrong}, {Fan}, {Finlator}, {Geballe}, {Gunn}, {Hennessy}, {Knapp}, {Leggett}, {Munn}, {Pier}, {Rockosi}, {Schneider}, {Strauss}, {Yanny}, {Brinkmann}, {Csabai}, {Hindsley}, {Kent}, {Lamb}, {Margon}, {McKay}, {Smith}, {Waddel}, {York}, \& {SDSS Collaboration}}]{2001AJ....122.2749I}
{Ivezi{\'c}}, {\v{Z}}., {Tabachnik}, S., {Rafikov}, R., {et~al.} 2001, \bibinfo{title}{{Solar System Objects Observed in the Sloan Digital Sky Survey Commissioning Data},} \aj, 122, 2749, \dodoi{10.1086/323452}

\bibitem[{{\v{Z}}. {Ivezi{\'c}} {et~al.}(2002){Ivezi{\'c}}, {Lupton}, {Juri{\'c}}, {Tabachnik}, {Quinn}, {Gunn}, {Knapp}, {Rockosi}, \& {Brinkmann}}]{2002AJ....124.2943I}
{Ivezi{\'c}}, {\v{Z}}., {Lupton}, R.~H., {Juri{\'c}}, M., {et~al.} 2002, \bibinfo{title}{{Color Confirmation of Asteroid Families},} \aj, 124, 2943, \dodoi{10.1086/344077}

\bibitem[{D. {Jewitt}(1991){Jewitt}}]{1991ASSL..167...19J}
{Jewitt}, D. 1991, \bibinfo{title}{{Cometary Photometry},} in Astrophysics and Space Science Library, Vol. 167, IAU Colloq. 116: Comets in the post-Halley era, ed. J.~{Newburn}, R.~L., M.~{Neugebauer}, \& J.~{Rahe}, 19, \dodoi{10.1007/978-94-011-3378-4_2}

\bibitem[{D. {Jewitt}(2009){Jewitt}}]{2009AJ....137.4296Jewitt}
{Jewitt}, D. 2009, \bibinfo{title}{{The Active Centaurs},} \aj, 137, 4296, \dodoi{10.1088/0004-6256/137/5/4296}

\bibitem[{D. {Jewitt}(2021){Jewitt}}]{2021AJ....161..261Jewitt}
{Jewitt}, D. 2021, \bibinfo{title}{{Systematics and Consequences of Comet Nucleus Outgassing Torques},} \aj, 161, 261, \dodoi{10.3847/1538-3881/abf09c}

\bibitem[{A. Johansen {et~al.}(2025)Johansen, Bannister, Dones, Jacobson, Singer, Volk, \& Womack}]{johansenchap2}
Johansen, A., Bannister, M., Dones, L., {et~al.} 2025, \bibinfo{title}{{Formation of Planetesimals in the Outer Solar System},} in Centaurs, ed. K.~{Volk}, M.~{Womack}, \& J.~{Steckloff} (IOP Publishing), 2--1\textemdash2--19, \dodoi{10.1088/2514-3433/ada267ch2}

\bibitem[{M. {Kaasalainen} \& J. {Torppa}(2001){Kaasalainen} \& {Torppa}}]{2001Icar..153...24K}
{Kaasalainen}, M., \& {Torppa}, J. 2001, \bibinfo{title}{{Optimization Methods for Asteroid Lightcurve Inversion. I. Shape Determination},} \icarus, 153, 24, \dodoi{10.1006/icar.2001.6673}

\bibitem[{M.~R. {Kirchoff} \& P. {Schenk}(2010){Kirchoff} \& {Schenk}}]{Kirchoff2010Icar}
{Kirchoff}, M.~R., \& {Schenk}, P. 2010, \bibinfo{title}{{Impact cratering records of the mid-sized, icy saturnian satellites},} \icarus, 206, 485, \dodoi{10.1016/j.icarus.2009.12.007}

\bibitem[{H. {Klahr} \& A. {Schreiber}(2020){Klahr} \& {Schreiber}}]{Klahr2020ApJ}
{Klahr}, H., \& {Schreiber}, A. 2020, \bibinfo{title}{{Turbulence Sets the Length Scale for Planetesimal Formation: Local 2D Simulations of Streaming Instability and Planetesimal Formation},} \apj, 901, 54, \dodoi{10.3847/1538-4357/abac58}

\bibitem[{R. Kokotanekova {et~al.}(2025)Kokotanekova, {Guilbert-Lepoutre}, Knight, \& Vincent}]{kokotanekovachap7}
Kokotanekova, R., {Guilbert-Lepoutre}, A., Knight, M.~M., \& Vincent, J.~B. 2025, \bibinfo{title}{{Evolutionary Processes in the Centaur Region},} in Centaurs, ed. K.~{Volk}, M.~{Womack}, \& J.~{Steckloff} (IOP Publishing), 7--1\textemdash7--29, \dodoi{10.1088/2514-3433/ada267ch7}

\bibitem[{P. {Lacerda}(2005){Lacerda}}]{2005PhDT........21Lacerda}
{Lacerda}, P. 2005, \bibinfo{title}{{The Shapes and Spins of Kuiper Belt Objects},} PhD thesis, Leiden Observatory

\bibitem[{P. {Lacerda}(2011){Lacerda}}]{2011AJ....142...90Lacerda}
{Lacerda}, P. 2011, \bibinfo{title}{{A Change in the Light Curve of Kuiper Belt Contact Binary (139775) 2001 QG$_{298}$},} \aj, 142, 90, \dodoi{10.1088/0004-6256/142/3/90}

\bibitem[{P. {Lacerda} {et~al.}(2014{\natexlab{a}}){Lacerda}, {McNeill}, \& {Peixinho}}]{2014MNRAS.437.3824Lacerda}
{Lacerda}, P., {McNeill}, A., \& {Peixinho}, N. 2014{\natexlab{a}}, \bibinfo{title}{{The unusual Kuiper belt object 2003 SQ$_{317}$},} \mnras, 437, 3824, \dodoi{10.1093/mnras/stt2180}

\bibitem[{P. {Lacerda} {et~al.}(2014{\natexlab{b}}){Lacerda}, {Fornasier}, {Lellouch}, {Kiss}, {Vilenius}, {Santos-Sanz}, {Rengel}, {M{\"u}ller}, {Stansberry}, {Duffard}, {Delsanti}, \& {Guilbert-Lepoutre}}]{2014ApJ...793L...2Lacerda}
{Lacerda}, P., {Fornasier}, S., {Lellouch}, E., {et~al.} 2014{\natexlab{b}}, \bibinfo{title}{{The Albedo-Color Diversity of Transneptunian Objects},} \apjl, 793, L2, \dodoi{10.1088/2041-8205/793/1/L2}

\bibitem[{P.~L. {Lamy} \& I. {Toth}(1995){Lamy} \& {Toth}}]{1995A&A...293L..43L}
{Lamy}, P.~L., \& {Toth}, I. 1995, \bibinfo{title}{{Direct detection of a cometary nucleus with the Hubble Space Telescope.},} \aap, 293, L43

\bibitem[{P.~L. {Lamy} {et~al.}(2004){Lamy}, {Toth}, {Fernandez}, \& {Weaver}}]{2004come.book..223L}
{Lamy}, P.~L., {Toth}, I., {Fernandez}, Y.~R., \& {Weaver}, H.~A. 2004, \bibinfo{title}{{The sizes, shapes, albedos, and colors of cometary nuclei},} in Comets II, ed. M.~C. {Festou}, H.~U. {Keller}, \& H.~A. {Weaver} (Univ. Arizona Press, Tucson, AZ), 223

\bibitem[{J. {Li} {et~al.}(2020){Li}, {Jewitt}, {Mutchler}, {Agarwal}, \& {Weaver}}]{2020AJ....159..209L}
{Li}, J., {Jewitt}, D., {Mutchler}, M., {Agarwal}, J., \& {Weaver}, H. 2020, \bibinfo{title}{{Hubble Space Telescope Search for Activity in High-perihelion Objects},} \aj, 159, 209, \dodoi{10.3847/1538-3881/ab7faf}

\bibitem[{J. {Licandro} {et~al.}(2012){Licandro}, {Hargrove}, {Kelley}, {Campins}, {Ziffer}, {Al{\'\i}-Lagoa}, {Fern{\'a}ndez}, \& {Rivkin}}]{Licandro2012A&A}
{Licandro}, J., {Hargrove}, K., {Kelley}, M., {et~al.} 2012, \bibinfo{title}{{5-14 {\ensuremath{\mu}}m Spitzer spectra of Themis family asteroids},} \aap, 537, A73, \dodoi{10.1051/0004-6361/201118142}

\bibitem[{C.~H. {Lineweaver} \& M. {Norman}(2010){Lineweaver} \& {Norman}}]{2010arXiv1004.1091Lineweaver}
{Lineweaver}, C.~H., \& {Norman}, M. 2010, \bibinfo{title}{{The Potato Radius: a Lower Minimum Size for Dwarf Planets},} arXiv e-prints, arXiv:1004.1091, \dodoi{10.48550/arXiv.1004.1091}

\bibitem[{C.~M. {Lisse} {et~al.}(1999){Lisse}, {Fern{\'a}ndez}, {Kundu}, {A'Hearn}, {Dayal}, {Deutsch}, {Fazio}, {Hora}, \& {Hoffmann}}]{1999Icar..140..189L}
{Lisse}, C.~M., {Fern{\'a}ndez}, Y.~R., {Kundu}, A., {et~al.} 1999, \bibinfo{title}{{The Nucleus of Comet Hyakutake (C/1996 B2)},} \icarus, 140, 189, \dodoi{10.1006/icar.1999.6131}

\bibitem[{P.-Y. {Liu} \& W.-H. {Ip}(2019){Liu} \& {Ip}}]{2019ApJ...880...71L}
{Liu}, P.-Y., \& {Ip}, W.-H. 2019, \bibinfo{title}{{An Investigation on the Origin of Centaurs{\textquoteright} Color-Inclination Relation},} \apj, 880, 71, \dodoi{10.3847/1538-4357/ab29eb}

\bibitem[{J. {Luu} \& D. {Jewitt}(1996){Luu} \& {Jewitt}}]{1996AJ....112.2310Luu}
{Luu}, J., \& {Jewitt}, D. 1996, \bibinfo{title}{{Color Diversity Among the Centaurs and Kuiper Belt Objects},} \aj, 112, 2310, \dodoi{10.1086/118184}

\bibitem[{J.~X. {Luu} \& D.~C. {Jewitt}(1990){Luu} \& {Jewitt}}]{1990AJ....100..913Luu}
{Luu}, J.~X., \& {Jewitt}, D.~C. 1990, \bibinfo{title}{{Cometary Activity in 2060 Chiron},} \aj, 100, 913, \dodoi{10.1086/115571}

\bibitem[{J.~X. {Luu} \& D.~C. {Jewitt}(1998){Luu} \& {Jewitt}}]{1998ApJ...494L.117Luu}
{Luu}, J.~X., \& {Jewitt}, D.~C. 1998, \bibinfo{title}{{Optical and Infrared Reflectance Spectrum of Kuiper Belt Object 1996 TL$_{66}$},} \apjl, 494, L117, \dodoi{10.1086/311172}

\bibitem[{A.~K. {Mainzer} {et~al.}(2019){Mainzer}, {Bauer}, {Cutri}, {Grav}, {Kramer}, {Masiero}, {Sonnett}, \& {Wright}}]{2019PDSS..251.....Mainzer}
{Mainzer}, A.~K., {Bauer}, J.~M., {Cutri}, R.~M., {et~al.} 2019, \bibinfo{title}{{NEOWISE Diameters and Albedos V2.0},} NASA Planetary Data System, \dodoi{10.26033/18S3-2Z54}

\bibitem[{A.~K. {Mainzer} {et~al.}(2023){Mainzer}, {Masiero}, {Abell}, {Bauer}, {Bottke}, {Buratti}, {Carey}, {Cotto-Figueroa}, {Cutri}, {Dahlen}, {Eisenhardt}, {Fernandez}, {Furfaro}, {Grav}, {Hoffman}, {Kelley}, {Kim}, {Kirkpatrick}, {Lawler}, {Lilly}, {Liu}, {Marocco}, {Marsh}, {Masci}, {McMurtry}, {Pourrahmani}, {Reinhart}, {Ressler}, {Satpathy}, {Schambeau}, {Sonnett}, {Spahr}, {Surace}, {Vaquero}, {Wright}, {Zengilowski}, \& {NEO Surveyor Mission Team}}]{2023PSJ.....4..224M}
{Mainzer}, A.~K., {Masiero}, J.~R., {Abell}, P.~A., {et~al.} 2023, \bibinfo{title}{{The Near-Earth Object Surveyor Mission},} \psj, 4, 224, \dodoi{10.3847/PSJ/ad0468}

\bibitem[{K. Mandt {et~al.}(2025)Mandt, Ivanova, H{arrington Pinto}, Roth, \& Seligman}]{mandtchap6}
Mandt, K., Ivanova, O., H{arrington Pinto}, O., Roth, N.~X., \& Seligman, D.~Z. 2025, \bibinfo{title}{{Volatiles},} in Centaurs, ed. K.~{Volk}, M.~{Womack}, \& J.~{Steckloff} (IOP Publishing), 6--1\textemdash6--34, \dodoi{10.1088/2514-3433/ada267ch6}

\bibitem[{R.~K. {Mann} {et~al.}(2007){Mann}, {Jewitt}, \& {Lacerda}}]{2007AJ....134.1133Mann}
{Mann}, R.~K., {Jewitt}, D., \& {Lacerda}, P. 2007, \bibinfo{title}{{Fraction of Contact Binary Trojan Asteroids},} \aj, 134, 1133, \dodoi{10.1086/520328}

\bibitem[{R. {Marschall} {et~al.}(2023){Marschall}, {Morbidelli}, {Bottke}, {Vokrouhlick{\'y}}, {Nesvorn{\'y}}, \& {Deienno}}]{Marschall2023ACM}
{Marschall}, R., {Morbidelli}, A., {Bottke}, W.~F., {et~al.} 2023, \bibinfo{title}{{Comets are fragments: What the Kuiper Belt size distribution tells us about its collisional evolution.},} in Asteroids, Comets, Meteors Conference 2023, 2470

\bibitem[{R. {Marschall} {et~al.}(2022){Marschall}, {Nesvorn{\'y}}, {Deienno}, {Wong}, {Levison}, \& {Bottke}}]{Marschall2022AJ}
{Marschall}, R., {Nesvorn{\'y}}, D., {Deienno}, R., {et~al.} 2022, \bibinfo{title}{{Implications for the Collisional Strength of Jupiter Trojans from the Eurybates Family},} \aj, 164, 167, \dodoi{10.3847/1538-3881/ac8d6b}

\bibitem[{W.~B. {McKinnon} {et~al.}(2020){McKinnon}, {Richardson}, {Marohnic}, {Keane}, {Grundy}, {Hamilton}, {Nesvorn{\'y}}, {Umurhan}, {Lauer}, {Singer}, {Stern}, {Weaver}, {Spencer}, {Buie}, {Moore}, {Kavelaars}, {Lisse}, {Mao}, {Parker}, {Porter}, {Showalter}, {Olkin}, {Cruikshank}, {Elliott}, {Gladstone}, {Parker}, {Verbiscer}, {Young}, \& {New Horizons Science Team}}]{2020Sci...367.6620M}
{McKinnon}, W.~B., {Richardson}, D.~C., {Marohnic}, J.~C., {et~al.} 2020, \bibinfo{title}{{The solar nebula origin of (486958) Arrokoth, a primordial contact binary in the Kuiper Belt},} Science, 367, aay6620, \dodoi{10.1126/science.aay6620}

\bibitem[{A. {McNeill} {et~al.}(2018){McNeill}, {Fitzsimmons}, {Jedicke}, {Lacerda}, {Lilly}, {Thompson}, {Trilling}, {DeMooij}, {Hooton}, \& {Watson}}]{2018AJ....156..282McNeill}
{McNeill}, A., {Fitzsimmons}, A., {Jedicke}, R., {et~al.} 2018, \bibinfo{title}{{Extreme Asteroids in the Pan-STARRS 1 Survey},} \aj, 156, 282, \dodoi{10.3847/1538-3881/aaeb8c}

\bibitem[{M. {Mommert} {et~al.}(2018){Mommert}, {Jedicke}, \& {Trilling}}]{2018AJ....155...74M}
{Mommert}, M., {Jedicke}, R., \& {Trilling}, D.~E. 2018, \bibinfo{title}{{An Investigation of the Ranges of Validity of Asteroid Thermal Models for Near-Earth Asteroid Observations},} \aj, 155, 74, \dodoi{10.3847/1538-3881/aaa23b}

\bibitem[{B.~E. {Morgado} {et~al.}(2021){Morgado}, {Sicardy}, {Braga-Ribas}, {Desmars}, {Gomes-J{\'u}nior}, {B{\'e}rard}, {Leiva}, {Ortiz}, {Vieira-Martins}, {Benedetti-Rossi}, {Santos-Sanz}, {Camargo}, {Duffard}, {Rommel}, {Assafin}, {Boufleur}, {Colas}, {Kretlow}, {Beisker}, {Sfair}, {Snodgrass}, {Morales}, {Fern{\'a}ndez-Valenzuela}, {Amaral}, {Amarante}, {Artola}, {Backes}, {Bath}, {Bouley}, {Buie}, {Cacella}, {Colazo}, {Colque}, {Dauvergne}, {Dominik}, {Emilio}, {Erickson}, {Evans}, {Fabrega-Polleri}, {Garcia-Lambas}, {Giacchini}, {Hanna}, {Herald}, {Hesler}, {Hinse}, {Jacques}, {Jehin}, {J{\o}rgensen}, {Kerr}, {Kouprianov}, {Levine}, {Linder}, {Maley}, {Machado}, {Maquet}, {Maury}, {Melia}, {Meza}, {Mondon}, {Moura}, {Newman}, {Payet}, {Pereira}, {Pollock}, {Poltronieri}, {Quispe-Huaynasi}, {Reichart}, {de Santana}, {Schneiter}, {Sieyra}, {Skottfelt}, {Soulier}, {Starck}, {Thierry}, {Torres}, {Trabuco}, {Unda-Sanzana}, {Yamashita}, {Winter}, {Zapata}, \& {Zuluaga}}]{2021A&A...652A.141Morgado}
{Morgado}, B.~E., {Sicardy}, B., {Braga-Ribas}, F., {et~al.} 2021, \bibinfo{title}{{Refined physical parameters for Chariklo's body and rings from stellar occultations observed between 2013 and 2020},} \aap, 652, A141, \dodoi{10.1051/0004-6361/202141543}

\bibitem[{S. {Mottola} {et~al.}(2024){Mottola}, {Britt}, {Brown}, {Buie}, {Noll}, \& {P{\"a}tzold}}]{2024SSRv..220...17M}
{Mottola}, S., {Britt}, D.~T., {Brown}, M.~E., {et~al.} 2024, \bibinfo{title}{{Shapes, Rotations, Photometric and Internal Properties of Jupiter Trojans},} \ssr, 220, 17, \dodoi{10.1007/s11214-024-01052-7}

\bibitem[{S. {Mottola} {et~al.}(2023){Mottola}, {Hellmich}, {Buie}, {Zangari}, {Stephens}, {Di Martino}, {Proffe}, {Marchi}, {Olkin}, \& {Levison}}]{2023PSJ.....4...18M}
{Mottola}, S., {Hellmich}, S., {Buie}, M.~W., {et~al.} 2023, \bibinfo{title}{{Shape Models of Lucy Targets (3548) Eurybates and (21900) Orus from Disk-integrated Photometry},} \psj, 4, 18, \dodoi{10.3847/PSJ/acaf79}

\bibitem[{T. {M{\"u}ller} {et~al.}(2020){M{\"u}ller}, {Lellouch}, \& {Fornasier}}]{2020tnss.book..153M}
{M{\"u}ller}, T., {Lellouch}, E., \& {Fornasier}, S. 2020, \bibinfo{title}{{Trans-Neptunian objects and Centaurs at thermal wavelengths},} in The Trans-Neptunian Solar System, ed. D.~{Prialnik}, M.~A. {Barucci}, \& L.~{Young}, 153--181, \dodoi{10.1016/B978-0-12-816490-7.00007-2}

\bibitem[{S.~A. {Myers} {et~al.}(2023){Myers}, {Howell}, {Magri}, {Vervack}, {Fern{\'a}ndez}, {Marshall}, \& {Taylor}}]{2023PSJ.....4....5M}
{Myers}, S.~A., {Howell}, E.~S., {Magri}, C., {et~al.} 2023, \bibinfo{title}{{Constraining the Limitations of NEATM-like Models: A Case Study with Near-Earth Asteroid (285263) 1998 QE2},} \psj, 4, 5, \dodoi{10.3847/PSJ/aca89d}

\bibitem[{F. {Namouni} \& M.~H.~M. {Morais}(2020){Namouni} \& {Morais}}]{2020MNRAS.494.2191N}
{Namouni}, F., \& {Morais}, M.~H.~M. 2020, \bibinfo{title}{{An interstellar origin for high-inclination Centaurs},} \mnras, 494, 2191, \dodoi{10.1093/mnras/staa712}

\bibitem[{D. {Nesvorn{\'y}}(2018){Nesvorn{\'y}}}]{Nesvorny2018ARA&A}
{Nesvorn{\'y}}, D. 2018, \bibinfo{title}{{Dynamical Evolution of the Early Solar System},} \araa, 56, 137, \dodoi{10.1146/annurev-astro-081817-052028}

\bibitem[{D. {Nesvorn{\'y}}(2021){Nesvorn{\'y}}}]{Nesvorny2021ApJL}
{Nesvorn{\'y}}, D. 2021, \bibinfo{title}{{Eccentric Early Migration of Neptune},} \apjl, 908, L47, \dodoi{10.3847/2041-8213/abe38f}

\bibitem[{D. {Nesvorn{\'y}} {et~al.}(2018){Nesvorn{\'y}}, {Vokrouhlick{\'y}}, {Bottke}, \& {Levison}}]{Nesvorny2018NatAs}
{Nesvorn{\'y}}, D., {Vokrouhlick{\'y}}, D., {Bottke}, W.~F., \& {Levison}, H.~F. 2018, \bibinfo{title}{{Evidence for very early migration of the Solar System planets from the Patroclus-Menoetius binary Jupiter Trojan},} Nature Astronomy, 2, 878, \dodoi{10.1038/s41550-018-0564-3}

\bibitem[{D. {Nesvorn{\'y}} {et~al.}(2020){Nesvorn{\'y}}, {Vokrouhlick{\'y}}, {Alexandersen}, {Bannister}, {Buchanan}, {Chen}, {Gladman}, {Gwyn}, {Kavelaars}, {Petit}, {Schwamb}, \& {Volk}}]{Nesvorny2020AJ}
{Nesvorn{\'y}}, D., {Vokrouhlick{\'y}}, D., {Alexandersen}, M., {et~al.} 2020, \bibinfo{title}{{OSSOS XX: The Meaning of Kuiper Belt Colors},} \aj, 160, 46, \dodoi{10.3847/1538-3881/ab98fb}

\bibitem[{K.~S. {Noll} {et~al.}(2008){Noll}, {Grundy}, {Stephens}, {Levison}, \& {Kern}}]{2008Icar..194..758N}
{Noll}, K.~S., {Grundy}, W.~M., {Stephens}, D.~C., {Levison}, H.~F., \& {Kern}, S.~D. 2008, \bibinfo{title}{{Evidence for two populations of classical transneptunian objects: The strong inclination dependence of classical binaries},} \icarus, 194, 758, \dodoi{10.1016/j.icarus.2007.10.022}

\bibitem[{K.~S. {Noll} {et~al.}(2006){Noll}, {Levison}, {Grundy}, \& {Stephens}}]{2006Icar..184..611Noll}
{Noll}, K.~S., {Levison}, H.~F., {Grundy}, W.~M., \& {Stephens}, D.~C. 2006, \bibinfo{title}{{Discovery of a binary Centaur},} \icarus, 184, 611, \dodoi{10.1016/j.icarus.2006.05.010}

\bibitem[{J.~L. {Ortiz} {et~al.}(2015){Ortiz}, {Duffard}, {Pinilla-Alonso}, {Alvarez-Candal}, {Santos-Sanz}, {Morales}, {Fern{\'a}ndez-Valenzuela}, {Licandro}, {Campo Bagatin}, \& {Thirouin}}]{2015A&A...576A..18Ortiz}
{Ortiz}, J.~L., {Duffard}, R., {Pinilla-Alonso}, N., {et~al.} 2015, \bibinfo{title}{{Possible ring material around centaur (2060) Chiron},} \aap, 576, A18, \dodoi{10.1051/0004-6361/201424461}

\bibitem[{N. {Peixinho} {et~al.}(2012){Peixinho}, {Delsanti}, {Guilbert-Lepoutre}, {Gafeira}, \& {Lacerda}}]{2012AandA...546A..86Peixinho}
{Peixinho}, N., {Delsanti}, A., {Guilbert-Lepoutre}, A., {Gafeira}, R., \& {Lacerda}, P. 2012, \bibinfo{title}{{The bimodal colors of Centaurs and small Kuiper belt objects},} \aap, 546, A86, \dodoi{10.1051/0004-6361/201219057}

\bibitem[{N. Peixinho {et~al.}(2025)Peixinho, Lilly, {Alvarez-Candal}, {Souza-Feliciano}, \& Seccull}]{peixinhochap5}
Peixinho, N., Lilly, E., {Alvarez-Candal}, A., {Souza-Feliciano}, A.~C., \& Seccull, T. 2025, \bibinfo{title}{{Surface Properties and Composition},} in Centaurs, ed. K.~{Volk}, M.~{Womack}, \& J.~{Steckloff} (IOP Publishing), 5--1\textemdash5--18, \dodoi{10.1088/2514-3433/ada267ch5}

\bibitem[{N. {Peixinho} {et~al.}(2020){Peixinho}, {Thirouin}, {Tegler}, {Di Sisto}, {Delsanti}, {Guilbert-Lepoutre}, \& {Bauer}}]{2020tnss.book..307Peixinho}
{Peixinho}, N., {Thirouin}, A., {Tegler}, S.~C., {et~al.} 2020, \bibinfo{title}{{From Centaurs to Comets - 40 years},} in The Trans-Neptunian Solar System, ed. D.~{Prialnik}, M.~A. {Barucci}, \& L.~{Young} (Elsevier), 307--329, \dodoi{10.1016/B978-0-12-816490-7.00014-X}

\bibitem[{S.~B. {Porter} {et~al.}(2024){Porter}, {Benecchi}, {Verbiscer}, {Grundy}, {Noll}, \& {Parker}}]{2024PSJ.....5..143P}
{Porter}, S.~B., {Benecchi}, S.~D., {Verbiscer}, A.~J., {et~al.} 2024, \bibinfo{title}{{Detection of Close Kuiper Belt Binaries with HST WFC3},} \psj, 5, 143, \dodoi{10.3847/PSJ/ad3f19}

\bibitem[{P. {Pravec} \& A.~W. {Harris}(2007){Pravec} \& {Harris}}]{2007Icar..190..250P}
{Pravec}, P., \& {Harris}, A.~W. 2007, \bibinfo{title}{{Binary asteroid population. 1. Angular momentum content},} \icarus, 190, 250, \dodoi{10.1016/j.icarus.2007.02.023}

\bibitem[{J.~D. {Ruprecht} {et~al.}(2015){Ruprecht}, {Bosh}, {Person}, {Bianco}, {Fulton}, {Gulbis}, {Bus}, \& {Zangari}}]{2015Icar..252..271Ruprecht}
{Ruprecht}, J.~D., {Bosh}, A.~S., {Person}, M.~J., {et~al.} 2015, \bibinfo{title}{{29 November 2011 stellar occultation by 2060 Chiron: Symmetric jet-like features},} \icarus, 252, 271, \dodoi{10.1016/j.icarus.2015.01.015}

\bibitem[{T.~K. {Safrit} {et~al.}(2021){Safrit}, {Steckloff}, {Bosh}, {Nesvorny}, {Walsh}, {Brasser}, \& {Minton}}]{2021PSJ.....2...14Safrit}
{Safrit}, T.~K., {Steckloff}, J.~K., {Bosh}, A.~S., {et~al.} 2021, \bibinfo{title}{{The Formation of Bilobate Comet Shapes through Sublimative Torques},} \psj, 2, 14, \dodoi{10.3847/PSJ/abc9c8}

\bibitem[{N.~H. {Samarasinha} \& B.~E.~A. {Mueller}(2013){Samarasinha} \& {Mueller}}]{2013ApJ...775L..10Samarasinha}
{Samarasinha}, N.~H., \& {Mueller}, B. E.~A. 2013, \bibinfo{title}{{Relating Changes in Cometary Rotation to Activity: Current Status and Applications to Comet C/2012 S1 (ISON)},} \apjl, 775, L10, \dodoi{10.1088/2041-8205/775/1/L10}

\bibitem[{U. {Sch{\"a}fer} {et~al.}(2017){Sch{\"a}fer}, {Yang}, \& {Johansen}}]{Schafer2017A&A}
{Sch{\"a}fer}, U., {Yang}, C.-C., \& {Johansen}, A. 2017, \bibinfo{title}{{Initial mass function of planetesimals formed by the streaming instability},} \aap, 597, A69, \dodoi{10.1051/0004-6361/201629561}

\bibitem[{P.~M. {Schenk} {et~al.}(2004){Schenk}, {Chapman}, {Zahnle}, \& {Moore}}]{Schenk2004jpsm}
{Schenk}, P.~M., {Chapman}, C.~R., {Zahnle}, K., \& {Moore}, J.~M. 2004, \bibinfo{title}{{Ages and interiors: the cratering record of the Galilean satellites},} in Jupiter. The Planet, Satellites and Magnetosphere, ed. F.~{Bagenal}, T.~E. {Dowling}, \& W.~B. {McKinnon}, Vol.~1 (Cambridge University Press), 427--456

\bibitem[{M.~E. {Schwamb} {et~al.}(2023){Schwamb}, {Jones}, {Yoachim}, {Volk}, {Dorsey}, {Opitom}, {Greenstreet}, {Lister}, {Snodgrass}, {Bolin}, {Inno}, {Bannister}, {Eggl}, {Solontoi}, {Kelley}, {Juri{\'c}}, {Lin}, {Ragozzine}, {Bernardinelli}, {Chesley}, {Daylan}, {{\v{D}}urech}, {Fraser}, {Granvik}, {Knight}, {Lisse}, {Malhotra}, {Oldroyd}, {Thirouin}, \& {Ye}}]{2023ApJS..266...22S}
{Schwamb}, M.~E., {Jones}, R.~L., {Yoachim}, P., {et~al.} 2023, \bibinfo{title}{{Tuning the Legacy Survey of Space and Time (LSST) Observing Strategy for Solar System Science},} \apjs, 266, 22, \dodoi{10.3847/1538-4365/acc173}

\bibitem[{S.~S. {Sheppard} \& D. {Jewitt}(2004){Sheppard} \& {Jewitt}}]{2004AJ....127.3023Sheppard}
{Sheppard}, S.~S., \& {Jewitt}, D. 2004, \bibinfo{title}{{Extreme Kuiper Belt Object 2001 QG$_{298}$ and the Fraction of Contact Binaries},} \aj, 127, 3023, \dodoi{10.1086/383558}

\bibitem[{M.~R. {Showalter} {et~al.}(2021){Showalter}, {Benecchi}, {Buie}, {Grundy}, {Keane}, {Lisse}, {Olkin}, {Porter}, {Robbins}, {Singer}, {Verbiscer}, {Weaver}, {Zangari}, {Hamilton}, {Kaufmann}, {Lauer}, {Mehoke}, {Mehoke}, {Spencer}, {Throop}, {Parker}, {Stern}, {New Horizons Geology}, \& Team}]{2021Icar..35614098Showalter}
{Showalter}, M.~R., {Benecchi}, S.~D., {Buie}, M.~W., {et~al.} 2021, \bibinfo{title}{{A statistical review of light curves and the prevalence of contact binaries in the Kuiper Belt},} \icarus, 356, 114098, \dodoi{10.1016/j.icarus.2020.114098}

\bibitem[{A. Sickafoose {et~al.}(2025)Sickafoose, {Giuliatti Winter}, Leiva, Olkin, Ragozzine, \& M.}]{sickafoosechap9}
Sickafoose, A., {Giuliatti Winter}, S.~M., Leiva, R., {et~al.} 2025, \bibinfo{title}{{The Near-Centaur Environment: Satellites, Rings, and Debris},} in Centaurs, ed. K.~{Volk}, M.~{Womack}, \& J.~{Steckloff} (IOP Publishing), 9--1\textemdash9--31, \dodoi{10.1088/2514-3433/ada267ch9}

\bibitem[{J.~B. {Simon} {et~al.}(2016){Simon}, {Armitage}, {Li}, \& {Youdin}}]{Simon2016ApJ}
{Simon}, J.~B., {Armitage}, P.~J., {Li}, R., \& {Youdin}, A.~N. 2016, \bibinfo{title}{{The Mass and Size Distribution of Planetesimals Formed by the Streaming Instability. I. The Role of Self-gravity},} \apj, 822, 55, \dodoi{10.3847/0004-637X/822/1/55}

\bibitem[{K.~N. {Singer} {et~al.}(2019){Singer}, {McKinnon}, {Gladman}, {Greenstreet}, {Bierhaus}, {Stern}, {Parker}, {Robbins}, {Schenk}, {Grundy}, {Bray}, {Beyer}, {Binzel}, {Weaver}, {Young}, {Spencer}, {Kavelaars}, {Moore}, {Zangari}, {Olkin}, {Lauer}, {Lisse}, {Ennico}, {New Horizons Geology}, Team, {New Horizons Surface Composition Science Theme Team}, \& {New Horizons Ralph and LORRI Teams}}]{Singer2019Sci}
{Singer}, K.~N., {McKinnon}, W.~B., {Gladman}, B., {et~al.} 2019, \bibinfo{title}{{Impact craters on Pluto and Charon indicate a deficit of small Kuiper belt objects},} Science, 363, 955, \dodoi{10.1126/science.aap8628}

\bibitem[{F. {Spoto} {et~al.}(2015){Spoto}, {Milani}, \& {Kne{\v{z}}evi{\'c}}}]{Spoto2015Icar}
{Spoto}, F., {Milani}, A., \& {Kne{\v{z}}evi{\'c}}, Z. 2015, \bibinfo{title}{{Asteroid family ages},} \icarus, 257, 275, \dodoi{10.1016/j.icarus.2015.04.041}

\bibitem[{J.~K. {Steckloff} \& N.~H. {Samarasinha}(2018){Steckloff} \& {Samarasinha}}]{2018Icar..312..172Steckloff}
{Steckloff}, J.~K., \& {Samarasinha}, N.~H. 2018, \bibinfo{title}{{The sublimative torques of Jupiter Family Comets and mass wasting events on their nuclei},} \icarus, 312, 172, \dodoi{10.1016/j.icarus.2018.04.031}

\bibitem[{S.~A. {Stern} {et~al.}(2019){Stern}, {Weaver}, {Spencer}, {Olkin}, {Gladstone}, {Grundy}, {Moore}, {Cruikshank}, {Elliott}, {McKinnon}, {Parker}, {Verbiscer}, {Young}, {Aguilar}, {Albers}, {Andert}, {Andrews}, {Bagenal}, {Banks}, {Bauer}, {Bauman}, {Bechtold}, {Beddingfield}, {Behrooz}, {Beisser}, {Benecchi}, {Bernardoni}, {Beyer}, {Bhaskaran}, {Bierson}, {Binzel}, {Birath}, {Bird}, {Boone}, {Bowman}, {Bray}, {Britt}, {Brown}, {Buckley}, {Buie}, {Buratti}, {Burke}, {Bushman}, {Carcich}, {Chaikin}, {Chavez}, {Cheng}, {Colwell}, {Conard}, {Conner}, {Conrad}, {Cook}, {Cooper}, {Custodio}, {Dalle Ore}, {Deboy}, {Dharmavaram}, {Dhingra}, {Dunn}, {Earle}, {Egan}, {Eisig}, {El-Maarry}, {Engelbrecht}, {Enke}, {Ercol}, {Fattig}, {Ferrell}, {Finley}, {Firer}, {Fischetti}, {Folkner}, {Fosbury}, {Fountain}, {Freeze}, {Gabasova}, {Glaze}, {Green}, {Griffith}, {Guo}, {Hahn}, {Hals}, {Hamilton}, {Hamilton}, {Hanley}, {Harch}, {Harmon}, {Hart}, {Hayes}, {Hersman}, {Hill}, {Hill}, {Hofgartner}, {Holdridge},
  {Hor{\'a}nyi}, {Hosadurga}, {Howard}, {Howett}, {Jaskulek}, {Jennings}, {Jensen}, {Jones}, {Kang}, {Katz}, {Kaufmann}, {Kavelaars}, {Keane}, {Keleher}, {Kinczyk}, {Kochte}, {Kollmann}, {Krimigis}, {Kruizinga}, {Kusnierkiewicz}, {Lahr}, {Lauer}, {Lawrence}, {Lee}, {Lessac-Chenen}, {Linscott}, {Lisse}, {Lunsford}, {Mages}, {Mallder}, {Martin}, {May}, {McComas}, {McNutt}, {Mehoke}, {Mehoke}, {Nelson}, {Nguyen}, {N{\'u}{\~n}ez}, {Ocampo}, {Owen}, {Oxton}, {Parker}, {P{\"a}tzold}, {Pelgrift}, {Pelletier}, {Pineau}, {Piquette}, {Porter}, {Protopapa}, {Quirico}, {Redfern}, {Regiec}, {Reitsema}, {Reuter}, {Richardson}, {Riedel}, {Ritterbush}, {Robbins}, {Rodgers}, {Rogers}, {Rose}, {Rosendall}, {Runyon}, {Ryschkewitsch}, {Saina}, {Salinas}, {Schenk}, {Scherrer}, {Schlei}, {Schmitt}, {Schultz}, {Schurr}, {Scipioni}, {Sepan}, {Shelton}, {Showalter}, {Simon}, {Singer}, {Stahlheber}, {Stanbridge}, {Stansberry}, {Steffl}, {Strobel}, {Stothoff}, {Stryk}, {Stuart}, {Summers}, {Tapley}, {Taylor}, {Taylor}, {Tedford},
  {Throop}, {Turner}, {Umurhan}, {Van Eck}, {Velez}, {Versteeg}, {Vincent}, {Webbert}, {Weidner}, {Weigle}, {Wendel}, {White}, {Whittenburg}, {Williams}, {Williams}, {Williams}, {Winters}, {Zangari}, \& {Zurbuchen}}]{2019Sci...364.9771S}
{Stern}, S.~A., {Weaver}, H.~A., {Spencer}, J.~R., {et~al.} 2019, \bibinfo{title}{{Initial results from the New Horizons exploration of 2014 MU$_{69}$, a small Kuiper Belt object},} Science, 364, aaw9771, \dodoi{10.1126/science.aaw9771}

\bibitem[{S.~C. {Tegler} {et~al.}(2016){Tegler}, {Romanishin}, {Consolmagno}, \& {J.}}]{2016AJ....152..210Tegler}
{Tegler}, S.~C., {Romanishin}, W., {Consolmagno}, G.~J., \& {J.}, S. 2016, \bibinfo{title}{{Two Color Populations of Kuiper Belt and Centaur Objects and the Smaller Orbital Inclinations of Red Centaur Objects},} \aj, 152, 210, \dodoi{10.3847/0004-6256/152/6/210}

\bibitem[{S.~C. {Tegler} {et~al.}(2005){Tegler}, {Romanishin}, {Consolmagno}, {Rall}, {Worhatch}, {Nelson}, \& {Weidenschilling}}]{2005Icar..175..390Tegler}
{Tegler}, S.~C., {Romanishin}, W., {Consolmagno}, G.~J., {et~al.} 2005, \bibinfo{title}{{The period of rotation, shape, density, and homogeneous surface color of the Centaur 5145 Pholus},} \icarus, 175, 390, \dodoi{10.1016/j.icarus.2004.12.011}

\bibitem[{P. {Th{\'e}bault}(2003){Th{\'e}bault}}]{2003EM&P...92..233Thebault}
{Th{\'e}bault}, P. 2003, \bibinfo{title}{{A Numerical Check of the Collisional Resurfacing Scenario},} Earth Moon and Planets, 92, 233, \dodoi{10.1023/B:MOON.0000031941.77871.c5}

\bibitem[{A. {Thirouin} \& S.~S. {Sheppard}(2018){Thirouin} \& {Sheppard}}]{2018AJ....155..248Thirouin}
{Thirouin}, A., \& {Sheppard}, S.~S. 2018, \bibinfo{title}{{The Plutino Population: An Abundance of Contact Binaries},} \aj, 155, 248, \dodoi{10.3847/1538-3881/aac0ff}

\bibitem[{D. {Vokrouhlick{\'y}} {et~al.}(2010){Vokrouhlick{\'y}}, {Nesvorn{\'y}}, {Bottke}, \& {Morbidelli}}]{Vokrouhlicky2010AJ}
{Vokrouhlick{\'y}}, D., {Nesvorn{\'y}}, D., {Bottke}, W.~F., \& {Morbidelli}, A. 2010, \bibinfo{title}{{Collisionally Born Family About 87 Sylvia},} \aj, 139, 2148, \dodoi{10.1088/0004-6256/139/6/2148}

\bibitem[{K. {Volk} \& R. {Malhotra}(2008){Volk} \& {Malhotra}}]{2008ApJ...687..714V}
{Volk}, K., \& {Malhotra}, R. 2008, \bibinfo{title}{{The Scattered Disk as the Source of the Jupiter Family Comets},} \apj, 687, 714, \dodoi{10.1086/591839}

\bibitem[{K. {Volk} {et~al.}(2025){Volk}, {Womack}, \& {Steckloff}}]{steckloffchap1}
{Volk}, K., {Womack}, M., \& {Steckloff}, J. 2025, \bibinfo{title}{{Introduction},} in Centaurs, ed. K.~{Volk}, M.~{Womack}, \& J.~{Steckloff} (IOP Publishing), 1--1\textemdash1--17, \dodoi{10.1088/2514-3433/ada267ch1}

\bibitem[{B.~D. {Warner} {et~al.}(2009){Warner}, {Harris}, \& {Pravec}}]{2009Icar..202..134Warner}
{Warner}, B.~D., {Harris}, A.~W., \& {Pravec}, P. 2009, \bibinfo{title}{{The asteroid lightcurve database},} \icarus, 202, 134, \dodoi{10.1016/j.icarus.2009.02.003}

\bibitem[{I. {Wong} \& M.~E. {Brown}(2015){Wong} \& {Brown}}]{Wong2015AJ}
{Wong}, I., \& {Brown}, M.~E. 2015, \bibinfo{title}{{The Color-Magnitude Distribution of Small Jupiter Trojans},} \aj, 150, 174, \dodoi{10.1088/0004-6256/150/6/174}

\bibitem[{I. {Wong} \& M.~E. {Brown}(2017){Wong} \& {Brown}}]{2017AJ....153..145Wong}
{Wong}, I., \& {Brown}, M.~E. 2017, \bibinfo{title}{{The Bimodal Color Distribution of Small Kuiper Belt Objects},} \aj, 153, 145, \dodoi{10.3847/1538-3881/aa60c3}

\bibitem[{E. {Wright} {et~al.}(2018){Wright}, {Mainzer}, {Masiero}, {Grav}, {Cutri}, \& {Bauer}}]{2018arXiv181101454W}
{Wright}, E., {Mainzer}, A., {Masiero}, J., {et~al.} 2018, \bibinfo{title}{{Response to ``An empirical examination of WISE/NEOWISE asteroid analysis and results''},} arXiv e-prints, arXiv:1811.01454, \dodoi{10.48550/arXiv.1811.01454}

\bibitem[{E.~L. {Wright}(2007){Wright}}]{2007astro.ph..3085W}
{Wright}, E.~L. 2007, \bibinfo{title}{{Comparing the NEATM with a Rotating, Cratered Thermophysical Asteroid Model},} arXiv e-prints, astro, \dodoi{10.48550/arXiv.astro-ph/0703085}

\bibitem[{F. {Yoshida} \& T. {Terai}(2017){Yoshida} \& {Terai}}]{Yoshida2017AJ}
{Yoshida}, F., \& {Terai}, T. 2017, \bibinfo{title}{{Small Jupiter Trojans Survey with the Subaru/Hyper Suprime-Cam},} \aj, 154, 71, \dodoi{10.3847/1538-3881/aa7d03}

\bibitem[{K. {Zahnle} {et~al.}(2003){Zahnle}, {Schenk}, {Levison}, \& {Dones}}]{Zahnle2003Icar}
{Zahnle}, K., {Schenk}, P., {Levison}, H., \& {Dones}, L. 2003, \bibinfo{title}{{Cratering rates in the outer Solar System},} \icarus, 163, 263, \dodoi{10.1016/S0019-1035(03)00048-4}

\end{thebibliography}
\bibliographystyle{aasjournalv7}



\end{document}